\documentclass{article}

\usepackage{arxiv}

\usepackage[utf8]{inputenc} 
\usepackage[T1]{fontenc}    
\usepackage{hyperref}       
\usepackage{url}            
\usepackage{booktabs}       
\usepackage{amsfonts}       
\usepackage{nicefrac}       
\usepackage{microtype}      
\usepackage{lipsum}		
\usepackage{graphicx}
\usepackage{natbib}
\usepackage{doi}
\usepackage{amsmath,amssymb,amsfonts}%
\usepackage{cleveref} 
\usepackage{natbib}
\usepackage{graphicx}%
\usepackage{multirow}%
\usepackage{amsthm}%
\usepackage{mathrsfs}%
\usepackage[figuresright]{rotating}%
\usepackage[title]{appendix}%
\usepackage{xcolor}%
\usepackage{textcomp}%
\usepackage{manyfoot}%
\usepackage{booktabs}%
\usepackage{algorithm}%
\usepackage{algorithmicx}%
\usepackage{algpseudocode}%
\usepackage{program}%
\usepackage{listings}%

\theoremstyle{thmstyleone}%
\newtheorem{remark}{Remark}%

\title{A finite element implementation of finite deformation surface and bulk poroelasticity}

\author{
    Jaemin ~Kim \\
	Sibley School of Mechanical and Aerospace Engineering \\
	Cornell University \\
	Ithaca, NY 14853, USA \\
	\And
    Ida ~Ang \\
	Sibley School of Mechanical and Aerospace Engineering \\
	Cornell University \\
	Ithaca, NY 14853, USA \\ 
	\And
    Francesco ~Ballarin \\
	Dipartimento di Matematica e Fisica “N. Tartaglia" \\
	Università Cattolica del Sacro Cuore \\
	via della Garzetta 48, Brescia 25133, Italy \\
	\And
    Chung-Yuen ~Hui \\
	Sibley School of Mechanical and Aerospace  \\
	Engineering, Cornell University \\
	Ithaca, NY 14853, USA \\
	Field of Theoretical and Applied Mechanics \\
	Cornell University \\
	Ithaca, NY 14853, USA \\
    Global Station for Soft Matter, GI-CoRE \\
    Hokkaido University \\
    Sapporo, Japan \\
	\And
    Nikolaos ~Bouklas\thanks{Corresponding author, nb589@cornell.edu} \\
	Sibley School of Mechanical and Aerospace  \\
	Engineering, Cornell University \\
	Ithaca, NY 14853, USA \\
    Field of Theoretical and Applied Mechanics \\
	Cornell University \\
	Ithaca, NY 14853, USA
}

\begin{document}
\maketitle

\begin{abstract}
We present a theoretical and computational model for the behavior of a porous solid undergoing two interdependent processes, the finite deformation of a solid and species migration through the solid, which are distinct in bulk and on surface. Nonlinear theories allow us to systematically study porous solids in a wide range of applications, such as drug delivery, biomaterial design, fundamental study of biomechanics and mechanobiology, and the design of sensors and actuators. As we aim to understand the physical phenomena at a smaller length scale towards comprehending the fundamental biological processes and the miniaturization of devices, the surface effect becomes more pertinent. Although existing methodologies provide the necessary tools to study coupled bulk effects for deformation and diffusion; however, very little is known about fully coupled bulk and surface poroelasticity at finite strain. Here we develop a thermodynamically consistent formulation for multiphysics processes of surface and bulk poroelasticity, specialized for soft hydrated solids, along with a corresponding finite element implementation. Our multiphysical approach captures the interplay between competing processes of finite deformation and species diffusion through the bulk and surface, and provides invaluable insight when surface effects are important.
\end{abstract}


\section{Introduction}
Poroelasticity is the theory that describes the coupling of deformations of a continuum body with diffusive flow of a solvent through the body itself due to its permeable/porous nature. The theory of linear poroelasticity as pioneered by Biot \citep{biot1955theory,biot1956theory,biot1962mechanics} has been applied to describe hydrogels and hydrated biological tissues \citep{hong2008theory,hong20153d,li2016designing}, but in many cases nonlinear variants of the theory of poroelasticity are necessary due to bulk deformation of finite deformations and nonlinear complex constitutive responses \citep{chester2010coupled,yoon2010poroelastic,li2012mechanics,bouklas2012swelling,bouklas2015nonlinear,ang2020effect}. These nonlinear theories have been highly relevant in the past few decades as the study of hydrated biomaterials, tissues and hydrogels at finite strains. Applications in drug delivery, tissue regeneration, biomaterials design, fundamental studies in biomechanics and mechanobiology, design of morphing structures as well as sensors and actuators, nonlinear theories of poroelasticity have provided a deep understanding of the rich observed material and structural responses. However, the majority of these studies have been in the macroscale, where surface effect can often be neglected as bulk energetic contributions dominate the response of the system. But, as we aim towards miniaturization of devices, and also aim to probe and fundamentally understand the effects that dominate the material response at smaller length scales surface effect become more relevant. When the length scales that are probed are sufficiently small, one can expect the coupling of elasticity and solvent diffusion to be significantly influenced by surface contributions toward all aspects of these multiphysical processes.

Following free energy-based phenomenological approaches for the bulk response of materials, similar approaches can can be developed to study the surface and interface response of materials and materials systems. The pioneering works of Gurtin \citep{gurtin1975continuum,GurtinMurdoch1978} established a systematic framework to derive kinematics and balance principles for surfaces that focused on the thermodynamics of purely elastic systems. Further, this framework was extended by Gurtin to account for thermal effects and mass transport particularized for surfaces \citep{gurtin2002interface,cermelli2005transport}. Many studies have followed since exploring the mechanics of surfaces and interfaces for continua involving purely mechanical as well as coupled processes \citep{dingreville2005surface,steinmann2008boundary,henann2014modeling,javili2014unified,liu2020modeling}.

The competition of bulk and surface energetics is length scale dependent. For elastic solids with shear modulus $\mathnormal{G}_{0}$ and surface energy $\widetilde{\gamma}$, a relevant length scale is the elastocapillary length $\ell=\widetilde{\gamma}/\mathnormal{G}_{0}$ \citep{style2017elastocapillarity}. Generally speaking, if the characteristic dimension is much larger than elastocapillary length ($\mathnormal{H}\gg\ell$), surface effects are negligible. On the other hand, if the characteristic dimension is much smaller than elastocapillary length ($\mathnormal{H}\ll\ell$), surface effects are dominant. Focusing on soft solids, work from \citet{style2017elastocapillarity} and \citet{bico2018elastocapillarity} focused on the implication of surface energetics on the mechanical behavior of this class of materials, looking at the subtle relationships between surface stresses and the bulk response. Surface and interface energetics have been further explored in the context of soft solids to study their implications to adhesion, fracture, instabilities as well as the design of composites. 

Specialized boundary value problems allow for the development of analytical solutions in this context \citep{bouklas2012swelling,ang2020effect}. However, finite element implementation that account for surface effects is necessary to tackle general problems. Notably, Javili and Steinmann \citep{javili2009finite,javili2010finite} developed the finite element framework for finite deformation accounting for surface energetics. Steinmann and co-authors \citep{mcbride2011geometrically,javili2013thermomechanics} also developed finite element schemes to study several classes of coupled problems, including mass transport and thermomechanics. These types of frameworks allowed the study of morphogenesis and wound healing in microtissue systems, where ``active'' surface and bulk effects were shown to emanate from the cellular activity of the extracellular matrix \citep{mailand2019surface,kim2020model,kim2023model}. Focused on drug delivery applications, Bouklas and co-authors \citep{ang2020effect} investigated the mechanical surface effect in the context of finite deformation poroelasticity, uncovering the importance of such effects on the transient response of swelling hydrogel microspheres of the porous solids coupled with solvent diffusion using a mixed finite element method. More recently, Dortdivanlioglu and co-authors \citep{rastogi2022modeling} proposed a computational formulation that allows the consideration of curvature-dependent surface energies using isogeometric analysis. 

There is an extensive body of work on the development of mixed finite element frameworks for linear poroelasticity, including stabilization approaches for the treatment of numerical instabilities that are known to arise due to the violation of the Ladyzhenskaya–Babuška–Brezzi (LBB) condition \citep{babuvska1971error,brezzi1974existence,murad1994stability,bathe2001inf}. For different variants of nonlinear poroelasticity, an array of approaches has been proposed \citep{hong2008theory,zhang2009finite,chester2010coupled,duda2010theory,bouklas2012swelling,lucantonio2013transient,chester2015finite,bouklas2015nonlinear,macminn2016large,ang2020effect,leronni2021modeling}. Although the existing methodologies provide the necessary tools to study coupled bulk effects for deformation and diffusion, to the best of the authors' knowledge an implementation of the finite element methods that fully couples bulk and surface poroelasticity at finite strain has not been previously presented. Motivated by the theoretical work of \citet{mcbride2011geometrically} that considered coupled thermomechanics  with species diffusion accounting for individual surface energetic contributions from elasticity, heat transfer and mass diffusion, here we aim to develop a robust and open-source finite element framework for the treatment of surface and bulk poroelasticity. In our previous work \citep{ang2020effect}, we considered the surface elasticity coupled with poroelasticity in bulk, not accounting for surface diffusion.

The ultimate goal of this paper is the development of a thermodynamically-consistent formulation for surface and bulk poroelasticity specialized for soft hydrated solids, along with the development of a corresponding finite element implementation. Focusing on soft solids, and more specifically hydrogels, we discover a set of numerical complications that are not evident when the general theoretical framework is presented. This is because of the nature of the coupling of surface and bulk poroelasticity, as well as the nearly incompressibility assumption that is utilized to model the elastomeric component of the hydrogel. The manuscript is organized as follows: In \Cref{sec:A_coupled_continuum_theory}, we describe the coupled theory, including kinematics, balance laws and the general form of the constitutive equations considering both bulk and surface energetics. In \Cref{sec:Specific_considerations_for_hydrogels}, the free energies are specialized for a hydrogel material. \Cref{sec:Mixed_finite_element_formulation} presents the finite element implementation including the weak form, normalization and solution procedure. \Cref{sec:Numerical_examples} elaborates on the numerical simulations for two three-dimensional (3D) boundary value problems corresponding to free-standing response of a cube-like structure as surface effects dominate, as well as the response under external loading highlighting unique contributions of surface effects, and the bulk and surface diffusion pathways. \Cref{sec:Conclusion} provides the conclusions and outlook for future work.

\section{A continuum multiphysics approach for bulk and surface poroelasticity}\label{sec:A_coupled_continuum_theory} 

In this section, we present a brief overview of key concept of differential geometry towards describing the kinematics, mechanical equilibrium and mass conservation, and obtain the general relationships for the constitutive relations for the bulk and surface responses. It is noteworthy that a surface quantity to be introduced is not necessarily the same as the bulk quantity evaluated on the surface. Let $\mathnormal{V}$ be a fixed reference configuration of a continuum body $\mathcal{B}$, where material points in the reference configuration are tracked through the vector $\textbf{X} \in \mathnormal{V}$.

\begin{remark}[Notation]\label{Rmk:Remark1}
     The dyad ($\otimes$) and dot ($\cdot$) denote the tensor and dot products, respectively. The double dot ($:$) denotes the double contraction over two second-order tensors; e.g. $\mathbf{A}:\mathbf{B}=\mathnormal{A}_{ij}\mathnormal{B}_{ij}$.  For convenience, we adopt the convention that other operations involving repeated indices will not be indicated by a specific symbol, specifically meaning for $\mathbf{v}=\mathbf{A}\mathbf{u}$ ($v_{i}=A_{ij}u_{j}$) and $\mathbf{A}=\mathbf{B}\mathbf{C}$ ($A_{ij}=B_{jk}C_{kj}$), we do not write $\mathbf{v}=\mathbf{A}\cdot\mathbf{u}$ and $\mathbf{A}=\mathbf{B}\cdot\mathbf{C}$, respectively. 
    Throughout the text, we use the Einstein summation convention: when an index variable appears twice in an expression, it implies summation of that variable over all values of the index. Latin indices take the value 1, 2, 3, and are used in variables describing the ambient, three-dimensional, Euclidean space. Greek indices take the value 1, 2, and are used in variables in the embedded, two-dimensional, surface. $\small\{\bullet\small\}$ and $\small\{\widetilde{\bullet}\small\}$ denote bulk and surface quantities, respectively, for a body occupying volume  $\mathnormal{V}$ bounded by outer surface denoted as $\mathnormal{S}$. The overhat $\{\widehat{\bullet}\}$ denotes normalized quantities, and the superposed dot $\{\dot\bullet\}$ denotes the material time derivative. We write $\mathbf{A}^{-1}=\left(\mathbf{A}\right)^{-1}$ and $\mathrm{tr}\left(\mathbf{A}\right)$ for the inverse and trace of a tensor $\mathbf{A}$.
\end{remark}

\subsection{Key concepts of differential geometry}\label{section:Geometry}

We briefly review the key concepts of differential geometry required to describe the kinematics of the bulk and surface. Further details on differential geometry can be found in books and papers, for example, Gurtin \citep{gurtin1975continuum}, Green \citep{green1992theoretical}, Steinmann \citep{steinmann2008boundary, javili2014unified} and Carmo \citep{do2016differential}.

We first focus on the bulk description. Let $\mathbf{X}$ be an arbitrary vector expressed in terms of the individual components of the curvilinear coordinate system $\xi^{i}$, i.e., $\mathbf{X}=\mathbf{X}(\xi^{1},\xi^{2},\xi^{3})$. We can always associate two new triplets of vectors ($\textbf{G}_{i}$ and $\textbf{G}^{i}$) with the general curvilinear coordinate $\xi^{i}$, by \citet{green1992theoretical}.

\begin{subequations}\label{generalized_coordinate_system}
\begin{align}
    \textbf{G}_{i}=\frac{\partial\mathbf{X}}{\partial\xi^{i}} \quad &\text{and} \quad  \textbf{G}^{i}=\frac{\partial\xi^{i}}{\partial\mathbf{X}}\\
    \textbf{I}=\delta^{i}_{j}\;\textbf{G}_{i}\otimes\textbf{G}^{j} \quad &\text{where} \quad \delta^{i}_{j}=\textbf{G}^{i}\cdot\textbf{G}_{j}
\end{align}
\end{subequations}
where $\textbf{G}_{i}$ and $\textbf{G}^{i}$ are referred to as covariant and contravariant basis vectors, respectively. $\textbf{I}$ and $\delta^{i}_{j}$ are the mixed-variant identity tensor and the mixed Kronecker delta, respectively. Note that $\delta^{i}_{j}=1$ for $i = j$, and $\delta^{i}_{j}=0$ for $i \neq j$ ($i,j=1,2,3$).

The covariant and contravariant basis vectors are not necessarily orthogonal to each other but linearly independent, requiring $(\textbf{G}_{1} \times \textbf{G}_{2}) \cdot \textbf{G}_{3} \neq 0$ (the vectors do not lie in a plane) \citep{holzapfel2000nonlinear}. The vectors $\textbf{G}_{i}$ and $\textbf{G}^{i}$ are connected to metric tensors $\textit{G}_{ij}$, $\textit{G}^{ij}$ by
\begin{subequations}\label{generalized_coordinate_system}
\begin{align}
    \textbf{G}_{i}=\textit{G}_{ij}\textbf{G}^{j} \quad \text{with} \quad \textit{G}_{ij}=\textit{G}_{ji}=\textbf{G}_{i}\cdot\textbf{G}_{j} \\
    \textbf{G}^{i}=\textit{G}^{ij}\textbf{G}_{j} \quad \text{with} \quad \textit{G}^{ij}=\textit{G}^{ji}=\textbf{G}^{i}\cdot\textbf{G}^{j}
\end{align}
\end{subequations}
where two tensors are inverse to each other, i.e., $\textit{G}^{ik}\textit{G}_{kj}=\delta^{i}_{j}$. Note that metric tensors with the identical indices represent the squares of the lengths; for example, $\textit{G}_{11}=\lvert\textbf{G}_{1}\rvert^{2}$, whereas the metric tensors with different indices represent the product of the lengths and the cosine of the angle $\theta$; for example, $\textit{G}_{12}=\lvert\textbf{G}_{1}\rvert\lvert\textbf{G}_{2}\rvert\cos\left(\theta(\textbf{G}_{1},\textbf{G}_{2})\right)$.

If a quantity $\small\{\bullet\small\}$ is a vector field defined throughout a volume $\textit{V}$ which is bounded by a closed surface $\textit{S}$, then bulk gradient and divergence operators in  3D general curvilinear coordinates are defined as follows:
\begin{subequations}
\begin{align}
    \boldsymbol{\nabla}_{ \textbf{X}}\small\{\bullet\small\}&=\frac{\partial\small\{\bullet\small\}}{\partial\xi^{i}}\otimes\textbf{G}^{i}\label{3D_gradient}\\
    \boldsymbol{\nabla}_{ \textbf{X}}\cdot\small\{\bullet\small\}&=\frac{\partial\small\{\bullet\small\}}{\partial\xi^{i}}\cdot\textbf{G}^{i}=\boldsymbol{\nabla}_{ \textbf{X}}\small\{\bullet\small\}:\textbf{I}\label{3D_divergence}
\end{align}
\end{subequations}
With the above equations, the divergence theorem is \citep{green1992theoretical}
\begin{equation}
    \int_V\boldsymbol{\nabla}_{ \textbf{X}}\cdot\,\small\{\bullet\small\}\,\mathrm{d}V=\int_S\small\{\bullet\small\}\cdot\textbf{N}\,\mathrm{d}S
    \label{3D_green_theorem}
\end{equation}
\noindent where $\textbf{N}$ is a unit outward normal vector on the surface $\mathnormal{S}$.

We now move our attention to the surface description. Let $\widetilde{\mathbf{X}}$ denote an arbitrary vector expressed in terms of the individual components of the general surface curvilinear coordinate system $\widetilde{\xi}{}^{\alpha}$, i.e., $\widetilde{\mathbf{X}}=\widetilde{\mathbf{X}}(\widetilde{\xi}{}^{1},\widetilde{\xi}{}^{2}$). We can always associate two new doublets of vectors ($\widetilde{\textbf{G}}_{\alpha}$ and $\widetilde{\textbf{G}}{}^{\alpha}$) with the general curvilinear coordinate $\widetilde{\xi}{}^{\alpha}$, by coordinate transformations \citep{green1992theoretical}.
\begin{subequations}\label{generalized_coordinate_system}
\begin{align}
    \widetilde{\textbf{G}}_{\alpha}=\frac{\partial\widetilde{\mathbf{X}}}{\partial\widetilde{\xi}{}^{\alpha}} \quad &\text{and} \quad  \widetilde{\textbf{G}}{}^{\alpha}=\frac{\partial\widetilde{\xi}{}^{\alpha}}{\partial\widetilde{\mathbf{X}}}\\
    \widetilde{\textbf{I}}=\delta{}^{\alpha}_{\beta}\;\widetilde{\textbf{G}}_{\alpha}\otimes\widetilde{\textbf{G}}{}{}^{\beta} \quad &\text{where} \quad \widetilde{\delta}{}^{\alpha}_{\beta}=\widetilde{\textbf{G}}{}^{\alpha}\cdot\widetilde{\textbf{G}}_{\beta}
\end{align}
\end{subequations}
where $\widetilde{\textbf{G}}_{\alpha}$ and $\widetilde{\textbf{G}}{}^{\alpha}$ are referred to as covariant and contravariant surface basis vectors, respectively. $\widetilde{\textbf{I}}$ and $\delta{}^{\alpha}_{\beta}$ are the mixed-variant surface identity tensor and the mixed Kronecker delta, respectively ($\alpha,\beta=1,2$). Note that the surface identity tensor also can be expressed by $\widetilde{\textbf{I}}=\textbf{I}-\textbf{N}\otimes\textbf{N}$.

The vectors $\widetilde{\textbf{G}}_{\alpha}$ and $\widetilde{\textbf{G}}{}^{\alpha}$ are connected with the geometrical characteristics
\begin{subequations}\label{generalized_coordinate_system}
\begin{align}
    \widetilde{\textbf{G}}_{\alpha}=\widetilde{\textit{G}}_{\,\alpha\beta}\widetilde{\textbf{G}}{}^{\beta} \quad \text{with} \quad \widetilde{\textit{G}}_{\,\alpha\beta}=\widetilde{\textit{G}}_{\,\beta\alpha}=\widetilde{\textbf{G}}_{\alpha}\cdot\widetilde{\textbf{G}}_{\beta}\\
    \widetilde{\textbf{G}}{}^{\alpha}=\widetilde{\textit{G}}{}^{\,\alpha\beta}\widetilde{\textbf{G}}_{\beta} \quad \text{with} \quad \widetilde{\textit{G}}{}^{\,\alpha\beta}=\widetilde{\textit{G}}{}^{\,\beta\alpha}=\widetilde{\textbf{G}}{}^{\alpha}\cdot\widetilde{\textbf{G}}{}^{\beta}
\end{align}
\end{subequations}
where $\widetilde{\textit{G}}_{\alpha\beta}$ and $\widetilde{\textit{G}}{}^{\,\alpha\beta}$ are surface metric tensors, and these two mappings are inverse to each other, i.e. $\widetilde{\textit{G}}{}^{\,\alpha\gamma}\widetilde{\textit{G}}_{\gamma\beta}=\widetilde{\delta}{}^{\alpha}_{\beta}$.

\begin{figure}[!ht]
    \centering
    \includegraphics[width=0.6\linewidth]{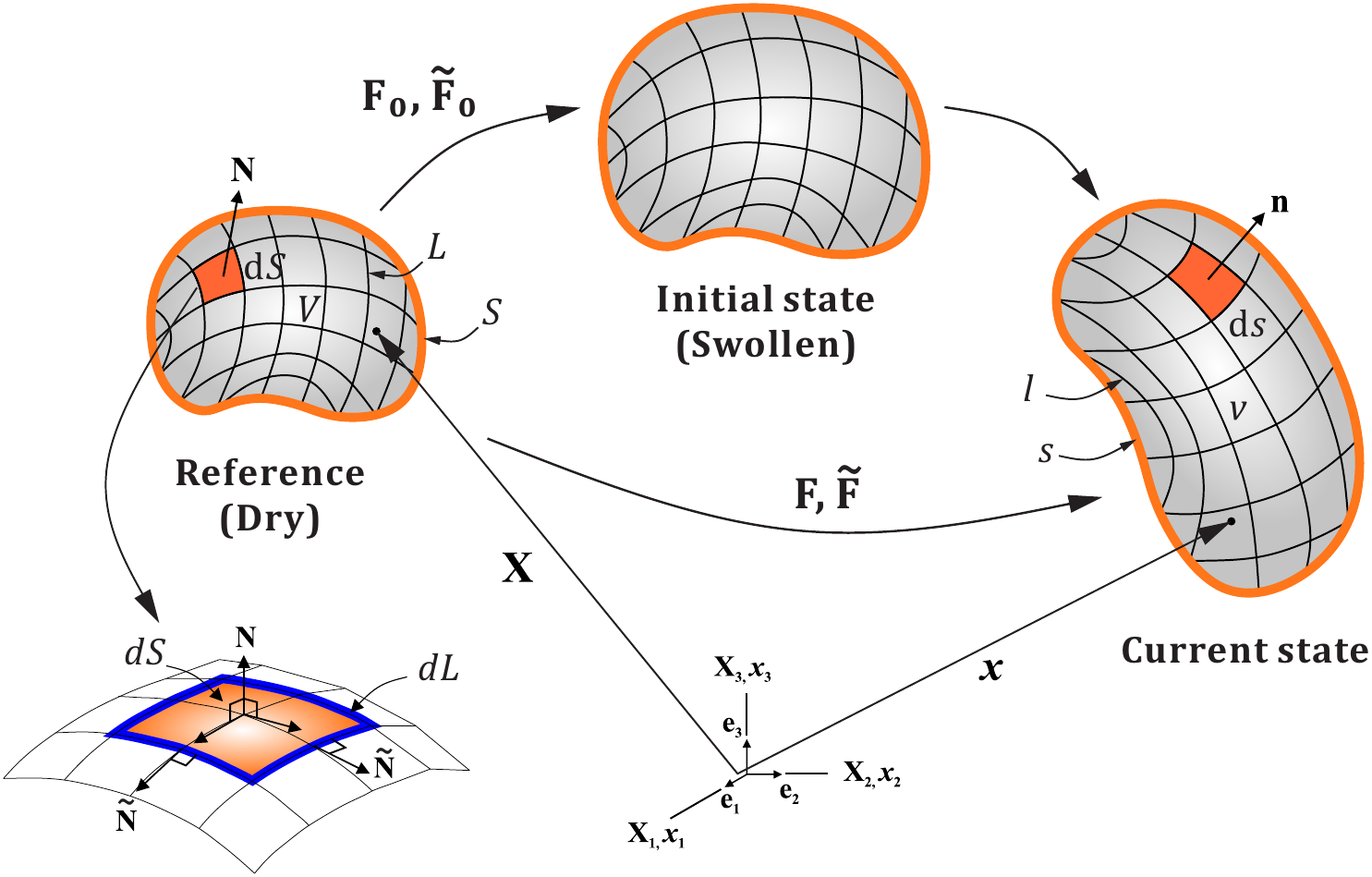}
    \caption{Schematic illustration of the reference, initial, and current state of a continuum body. The initial state is assumed to be isotropically scaled from the reference state. The reference volume and surface, and boundary are denoted by $\mathnormal{V}$ and $\mathnormal{S}$, and $\mathnormal{L}$ respectively. The normal vector to the surface in the reference and current configuration ($\mathbf{N}$ and $\mathbf{n}$) and the bi-normal vector to the boundary ($\widetilde{\mathbf{N}}$) are shown, where the over-tilde indicates surface quantities.}
    \label{fig:Fig1}
\end{figure}

If a quantity $\small\{\widetilde{\bullet}\small\}$ is a surface vector field defined throughout a surface $\textit{S}$, which is bounded by a closed curve $\textit{L}$, then the surface gradient and divergence operators in a 2D general curvilinear coordinates are defined as follows:
\begin{subequations}
\begin{align}
    \widetilde{\boldsymbol{\nabla}}_{ \widetilde{\textbf{X}}}\small\{\widetilde{\bullet}\small\}&=\frac{\partial\small\{\widetilde{\bullet}\small\}}{\partial\widetilde{\xi}{}^{\alpha}}\otimes\widetilde{\textbf{G}}{}^{\alpha}\label{3D_gradient}\\
    \widetilde{\boldsymbol{\nabla}}_{ \widetilde{\textbf{X}}}\cdot\small\{\widetilde{\bullet}\small\}&=\frac{\partial\small\{\widetilde{\bullet}\small\}}{\partial\widetilde{\xi}{}^{\alpha}}\cdot\widetilde{\textbf{G}}{}^{\alpha}=\widetilde{\boldsymbol{\nabla}}_{ \widetilde{\textbf{X}}}\small\{\widetilde{\bullet}\small\}:\widetilde{\textbf{I}}\label{3D_divergence}
\end{align}
\end{subequations}
With the above equations, we introduce the surface divergence theorem as follows \citep{steinmann2008boundary}:
\begin{equation}
    \int_{\mathnormal{S}}\widetilde{\boldsymbol{\nabla}}_{ \widetilde{\textbf{X}}}\cdot\{\widetilde{\bullet}\}\,\text{d}S
    =\int_{\mathnormal{L}}\small\{\widetilde{\bullet}\small\}\cdot\widetilde{\textbf{N}}\,\text{d}\mathnormal{L}
    -\int_{\mathnormal{S}}\widetilde{\kappa}\small\{\widetilde{\bullet}\small\}\cdot\textbf{N}\,\text{d}S    \label{2D_green_theorem}
\end{equation}
where $\widetilde{\textbf{N}}$ is the unit outward  bi-normal vector to the boundary curve $\mathnormal{L}$ (see \Cref{fig:Fig1}), and $\widetilde{\kappa}=-\widetilde{\boldsymbol{\nabla}}_{ \widetilde{\textbf{X}}}\cdot\textbf{N}$ is the total curvature (twice the mean surface curvature) \citep{gurtin1975continuum}. 

\subsection{Kinematics}
We use the notation $\chi : \mathnormal{V} \rightarrow \mathbb{R}^3$ for the deformation of body $\mathcal{B}$. A motion $\chi$ is the vector field of the mapping $\textit{\textbf{x}} = \chi(\textbf{X},\mathnormal{t})$, of a material point in the reference configuration $\textbf{X} \in \mathnormal{V}$ to a position in the deformed configuration $\textit{\textbf{x}} \in \mathnormal{v}$. The kinematics of a typical particle are described by the displacement vector field in the spatial description, $\textbf{u}(\textbf{X},t)=\textit{\textbf{x}}\,(\textbf{X},t)-\textbf{X}$. The kinematics of an infinitesimal bulk element are described by
\begin{subequations}\label{deformation_gradient_bulk}
\begin{align}
    \textbf{F}(\textbf{X},\textit{t})&=\frac{\partial\chi(\textbf{X},\textit{t})}{\partial\textbf{X}}=\boldsymbol{\nabla}_{ \textbf{X}}\,\textit{\textbf{x}}\,(\mathbf{X},t)\label{forward_deformation_gradient_bulk}\\
    \textbf{F}^{-1}(\textit{\textbf{x}},\textit{t})&=\frac{\partial\chi^{-1}(\textit{\textbf{x}},\textit{t})}{\partial\textit{\textbf{x}}}=\boldsymbol{\nabla}_{ \textit{\textbf{x}}}\,\textbf{X}(\textit{\textbf{x}},\textit{t})\label{inverse_deformation_gradient_bulk}
\end{align}
\end{subequations}
where $\textbf{F}(\textbf{X},\textit{t})$ and $\textbf{F}^{-1}(\textit{\textbf{x}},\textit{t})$ are the deformation gradient and inverse deformation gradient, respectively. Note that $\mathnormal{J}(\textbf{X},\textit{t})=\textrm{d}\textit{v}/\textrm{d}\textit{V}=\text{det} \, \textbf{F}(\textbf{X},\textit{t})>0$ is the Jacobian determinant of the deformation gradient defining the ratio of a volume element between the material and spatial configuration.

Assuming conformity, the surface displacement $\widetilde{\textbf{u}}(\widetilde{\textbf{X}},t)$ can be determined by $\mathbf{u}(\mathbf{X},t)\vert_{\mathnormal{S}} = \widetilde{\textbf{u}}(\widetilde{\textbf{X}},t)$. The motion of an arbitrary differential vector element $\mathrm{d}\mathbf{X}$ can be mapped by the deformation gradient $\textbf{F}$ to a vector $\mathrm{d}\mathbf{x}$ in the deformed configuration. However, a unit normal vector $\textbf{N}$ in the material configuration cannot be transformed into a unit normal vector $\textbf{n}$ in the spatial configuration using solely the deformation gradient \citep{holzapfel2000nonlinear,steinmann2008boundary}. This motivates us to follow the kinematics of an infinitesimal surface element \citep{kim2020model,kim2023model}. 
\begin{subequations}\label{deformation_gradient_surf}
\begin{align}
    \widetilde{\textbf{F}}(\widetilde{\textbf{X}},\textit{t})&
    =\frac{\partial\chi(\widetilde{\textbf{X}},\textit{t})}{\partial\widetilde{\textbf{X}}}\cdot\widetilde{\mathbf{I}}
    =\widetilde{\boldsymbol{\nabla}}_{ \widetilde{\textbf{X}}}\,\widetilde{\textit{\textbf{x}}}\,(\widetilde{\mathbf{X}},t)\label{forward_deformation_gradient_surf}\\
    \widetilde{\textbf{F}}^{-1}(\widetilde{\textit{\textbf{x}}},\textit{t})&
    =\frac{\partial\chi^{-1}(\widetilde{\textit{\textbf{x}}},\textit{t})}{\partial\widetilde{\textit{\textbf{x}}}}\cdot\widetilde{\textit{\textbf{i}}}
    =\widetilde{\boldsymbol{\nabla}}_{\widetilde{ \textit{\textbf{x}}}}\,\widetilde{\textbf{X}}(\widetilde{\textit{\textbf{x}}},\textit{t})\label{inverse_deformation_gradient_surf}
\end{align}
\end{subequations}
where $\widetilde{\textbf{F}} (\widetilde{\textbf{X}}, \textit{t})$ and $\widetilde{\textbf{F}}^{-1} (\widetilde{\textit{\textbf{x}}}, \textit{t})$ are the surface deformation gradient and inverse surface deformation gradient. Note $\widetilde{\textbf{F}}$ is a mapping from the surface (2D) to the bulk (3D), so it is rank-deficient (it has rank 2, whereas full-rank would be 3). The $\widetilde{\textbf{I}}=\textbf{I}-\textbf{N}\otimes\textbf{N}$ and $\widetilde{\textit{\textbf{i}}}=\textit{\textbf{i}}-\textbf{n}\otimes\textbf{n}$ are the mixed surface unit tensors with the outward unit normal vectors $\textbf{N}$ and $\textbf{n}$, where $\widetilde{\mathbf{I}}$ and $\widetilde{\textit{\textbf{i}}}$ act as a surface (idempotent) projection tensors in material and spatial configurations, respectively. Note that $\widetilde{\mathnormal{J}}(\widetilde{\textbf{X}},\textit{t})=\textrm{d}\textit{a}/\textrm{d}\textit{A}=\text{det} \, \widetilde{\textbf{F}}(\widetilde{\textbf{X}},\textit{t})>0$ is the Jacobian determinant of the surface deformation gradient defining the area ratio of a surface element between material and spatial configuration.

We introduce  $\mathbf{C}$ and $\widetilde{\mathbf{C}}$, the right Cauchy-Green tensors in the bulk and on the surface, respectively, as
\begin{subequations}
\begin{align}
    \mathbf{C} &= \mathbf{F}^{\mathrm{T}}\mathbf{F} \\ 
    \widetilde{\textbf{C}} &= \widetilde{\textbf{F}}^{\text{T}} \, \widetilde{\textbf{F}}
\end{align}
\end{subequations}
and $\mathnormal{I}_{1} = \text{tr}(\mathbf{C})$ is the first principal invariant. The detailed derivation for the surface kinematics can be found in \citet{green1992theoretical,steinmann2008boundary,do2016differential}. Note that we cannot perform the inverse of the surface right Cauchy-Green tensor due to its rank deficiency. Nevertheless, we can still obtain its inverse form in the generalized sense,
\begin{equation}
   \widetilde{\mathbf{C}}^{-1} = \widetilde{\mathbf{I}}\mathbf{C}^{-1}\widetilde{\mathbf{I}}
\end{equation}
which will be utilized in the forthcoming developments for defining the surface kinetic law.

\subsection{Mechanical equilibrium}
Mechanical equilibrium is assumed to be maintained at all times during the motion. The strong form of the corresponding governing equation is
\begin{subequations}\label{eqn:Equilibrium}
\begin{alignat}{2}
    \boldsymbol{\nabla}_{\textbf{X}}\cdot\mathbf{P} + \mathbf{B} &= 0 \quad  &&\text{in} \quad \mathnormal{V} \label{eqn:EquilibriumBulk} \\
    \mathbf{P}\mathbf{N} - \widetilde{\boldsymbol{\nabla}}_{\widetilde{\textbf{X}}}\cdot\widetilde{\mathbf{P}} &= \mathbf{T} \quad  &&\text{on} \quad \mathnormal{S}_{\mathrm{T}} \label{eqn:EquilibriumSurf}\\
    \mathbf{u} &= \mathbf{u}_{p} \quad &&\text{on} \quad \mathnormal{S}_{\mathrm{u}} \label{eqn:EqulibriumDirchletBC}\\
    [[\widetilde{\mathbf{P}}\widetilde{\mathbf{N}}]] &= 0 \quad  &&\text{on} \quad \mathnormal{L} \label{eqn:EqulibriumNeumannBC}
\end{alignat}
\end{subequations}
where $\mathbf{P}$ and $\widetilde{\mathbf{P}}$ are the first Piola-Kirchoff stresses in bulk and on surface, $\mathbf{B}$ and $\mathbf{T}$ are the body force and the traction vector, and $\mathbf{u}_{p}$ is the prescribed displacement. Note that a Neumann-type boundary condition is also defined on boundary curves that $[[\bullet]]$ indicates summation over surfaces intersecting on boundary curves \citep{steinmann2008boundary}.

\subsection{Species conservation}
Through species (or mass) conservation, the strong form for the corresponding governing equation is \citep{mcbride2011geometrically}:
\begin{subequations}\label{eqn:ConservationLaw}
\begin{alignat}{2}
    \dot{\mathnormal{C}} + \boldsymbol{\nabla}_{\textbf{X}} \cdot \mathbf{J} &= \mathnormal{r}
    \quad &&\text{in} \quad \mathnormal{V} \label{eqn:ConservationLawBulk}\\
    \dot{\widetilde{\mathnormal{C}}} + \widetilde{\boldsymbol{\nabla}}_{\textbf{X}} \cdot \widetilde{\mathbf{J}}-\mathbf{J}\cdot\mathbf{N} &= \mathnormal{i}
    \quad &&\text{on} \quad \mathnormal{S}_{\widetilde{\mathnormal{C}}} \label{eqn:ConservationLawSurf}\\
    \widetilde{\mathbf{J}} &= \widetilde{\mathbf{J}}_{p} \quad &&\text{on} \quad \mathnormal{S}_{\widetilde{\mathbf{J}}} \label{eqn:ConservationLawDirchletBC} \\
    [[\widetilde{\mathbf{J}} \cdot \widetilde{\mathbf{N}}]] &= 0
    \quad &&\text{on} \quad \mathnormal{L} \label{eqn:ConservationLawNeumannBC}
\end{alignat}
\end{subequations}
where $\mathnormal{C}$ is the bulk nominal concentration (the number of species per unit reference volume), $\widetilde{\mathnormal{C}}$ is the surface nominal concentration (the number of species per unit reference area), $\mathbf{J}$ is the bulk nominal flux (the number of species per unit time per unit area), and $\widetilde{\mathbf{J}}$ is the surface nominal flux (the number of species per unit time per unit length). The $\mathnormal{r}$ and $\mathnormal{i}$ are the source/sink terms for the number of species injected into the reference volume and area per unit time, and $\widetilde{\mathbf{J}}_{p}$ is the prescribed surface flux. \Cref{eqn:ConservationLawBulk} is the standard species conservation equation prescribed for the bulk in the reference configuration.  \Cref{eqn:ConservationLawSurf} describes the species balance on the surface, where the third term on the left hand is similar to the source $\mathnormal{i}$, but describes the outward normal flux from the bulk to the surface; note that species can get to the surface either from the exterior (environment) or from the interior (bulk). Additionally, \Cref{eqn:ConservationLaw} are supplemented with initial conditions,
\begin{subequations}\label{eqn:InitialCondition}
\begin{align}
    \mathnormal{C}(\mathbf{X},\mathnormal{t}=0) &= \mathnormal{C}_{0}\\
    \widetilde{\mathnormal{C}}(\widetilde{\mathbf{X}},\mathnormal{t}=0) &= \widetilde{\mathnormal{C}}_{0} 
\end{align}
\end{subequations}
where $\mathnormal{C}_{0}$ and $\widetilde{\mathnormal{C}}_{0}$ are the initial species concentration in the bulk and on the surface at time $t=0$. Note that we assume homogeneous initial conditions; this choice will become apparent as we discuss the general solution process for the corresponding boundary value problems in Section \ref{sec:Numerical_examples}.

\subsection{Thermodynamic considerations}\label{subsec:Thermodynamic considerations}
In addition to the free energy density in the bulk, we consider the free energy density on the surface. We will denote the bulk and surface free energy densities as
\begin{equation}\label{eqn:StrainEnergyFunction}
    \Psi(\mathbf{F},\mathnormal{C}) \quad \text{and} \quad \widetilde{\Psi}(\widetilde{\mathbf{F}},\widetilde{\mathnormal{C}})
\end{equation}
where we assume that they
are functions of the deformation gradient and nominal concentration in the bulk and on the surface, respectively.

Considering a system that includes an elastic and porous solid component coupled with species that are free to migrate in the porous network, the rate of change of the system's free energy $\mathcal{G}$ has to account for several effects  \citep{holzapfel2000nonlinear,gurtin2010mechanics,hong2008theory,bouklas2015nonlinear,ang2020effect}. This can be expressed as
\begin{equation} \label{eqn:FreeEnergyOfSystem}
    \dot{\mathcal{G}}
    = \int_{\mathnormal{V}} \dot{\Psi} \,\mathrm{d}V
    + \int_{\mathnormal{S}} \dot{\widetilde{\Psi}} \,\mathrm{d}S
    - \int_{\mathnormal{V}} \mathbf{B}\dot{\mathbf{x}} \,\mathrm{d}V
    - \int_{\mathnormal{S}} \mathbf{T}\dot{\mathbf{x}} \,\mathrm{d}S 
    - \int_{\mathnormal{V}} \mu\mathnormal{r} \,\mathrm{d}V
    - \int_{\mathnormal{S}} \widetilde{\mu}\mathnormal{i} \,\mathrm{d}S
\end{equation}
where the third and fourth terms are the rate of mechanical work by the body force $\mathbf{B}$ and traction vector $\mathbf{T}$, and the fifth and sixth terms are the rate of chemical work by the bulk chemical potential $\mu$ and the surface chemical potential $\widetilde{\mu}$. Note that thermodynamics dictate that the free energy of the system should not increase, i.e., $\dot{\mathcal{G}}\leq 0$.

Substituting \Cref{eqn:Equilibrium} and \Cref{eqn:ConservationLaw} into \Cref{eqn:FreeEnergyOfSystem}, the rate of change of the free energy of the system can be expressed as follows,
\begin{align}
    \dot{\mathcal{G}}
    &= \int_{\mathnormal{V}} \dot{\Psi} \,\mathrm{d}V
    +\int_{\mathnormal{S}} \dot{\widetilde{\Psi}} \,\mathrm{d}S
    - \int_{\mathnormal{V}} \mathbf{P}:\dot{\mathbf{F}} \,\mathrm{d}V 
    - \int_{\mathnormal{S}}  \widetilde{\mathbf{P}}:\dot{\widetilde{\mathbf{F}}} \,\mathrm{d}S
    \nonumber\\
    &- \int_{\mathnormal{V}} \mu\dot{\mathnormal{C}} \,\mathrm{d}V 
    - \int_{\mathnormal{S}} \widetilde{\mu}\dot{\widetilde{\mathnormal{C}}} \,\mathrm{d}S
    + \int_{\mathnormal{S}} \left(\widetilde{\mu}-\mu\right) \mathbf{J}\cdot\mathbf{N} \,\mathrm{d}S
    \nonumber\\
    &+ \int_{\mathnormal{V}} \mathbf{J}\cdot\boldsymbol{\nabla}_{\mathbf{X}}\mu \,\mathrm{d}V
    + \int_{\mathnormal{S}} \widetilde{\mathbf{J}}\cdot\widetilde{\boldsymbol{\nabla}}_{\widetilde{\textbf{X}}}\widetilde{\mu} \,\mathrm{d}S
    \leq 0 \label{eqn:DissipationInequality}
\end{align}
Using the chain-rule, the rate of bulk and surface free energy densities can be expressed as
\begin{subequations}\label{eqn:StrainEnergyChainRule}
\begin{align}
    \dot{\Psi} 
    &= \frac{\partial\Psi}{\partial\mathbf{F}}:\dot{\mathbf{F}} + \frac{\partial\Psi}{\partial\mathnormal{C}}\dot{\mathnormal{C}}
    \\
    \dot{\widetilde{\Psi}}
    &= \frac{\partial\widetilde{\Psi}}{\partial\widetilde{\mathbf{F}}}:\dot{\widetilde{\mathbf{F}}} + \frac{\partial\widetilde{\Psi}}{\partial\widetilde{\mathnormal{C}}}\dot{\widetilde{\mathnormal{C}}}
\end{align}
\end{subequations}
By substituting \Cref{eqn:StrainEnergyChainRule} into \Cref{eqn:DissipationInequality}, and rearranging terms yields
\begin{align}
    \dot{\mathcal{G}}
    &= \int_{\mathnormal{V}} \left(\frac{\partial\Psi}{\partial\mathbf{F}}-\mathbf{P}\right):\dot{\mathbf{F}} \,\mathrm{d}V
    + \int_{\mathnormal{S}} \left(\frac{\partial\widetilde{\Psi}}{\partial\widetilde{\mathbf{F}}}-\widetilde{\mathbf{P}}\right):\dot{\widetilde{\mathbf{F}}} \,\mathrm{d}S 
    \nonumber\\
    &+ \int_{\mathnormal{V}} \left(\frac{\partial\Psi}{\partial\mathnormal{C}}-\mu\right)\dot{\mathnormal{C}} \,\mathrm{d}V 
    + \int_{\mathnormal{S}} \left(\frac{\partial\widetilde{\Psi}}{\partial\widetilde{\mathnormal{C}}}-\widetilde{\mu}\right)\dot{\widetilde{\mathnormal{C}}} \,\mathrm{d}S \nonumber\\
    &+ \int_{\mathnormal{S}} \left(\widetilde{\mu}-\mu\right) \mathbf{J}\cdot\mathbf{N} \,\mathrm{d}S 
    + \int_{\mathnormal{V}} \mathbf{J}\cdot\boldsymbol{\nabla}_{\mathbf{X}}\mu \,\mathrm{d}V
    + \int_{\mathnormal{S}} \widetilde{\mathbf{J}}\cdot\widetilde{\boldsymbol{\nabla}}_{\widetilde{\textbf{X}}}\widetilde{\mu} \,\mathrm{d}S
    \leq 0 \label{eqn:DissipationInequalityChainRule}
\end{align}
where each integral represents a distinct mechanism of energy dissipation, associated with mechanical and chemical works. The inequality must hold at every point of the continuum body and at all times during a thermodynamic process. To satisfy the constraint, the Coleman-Noll procedure \citep{holzapfel2000nonlinear} states that each integrand in \Cref{eqn:DissipationInequalityChainRule} to be either negative or equal to zero.

From the first four terms in \Cref{eqn:DissipationInequalityChainRule} we obtain the following constitutive relations
\begin{subequations}\label{eqn:ConstitutiveRelation}
\begin{alignat}{3}
    \mathbf{P} &= \frac{\partial\Psi(\mathbf{F},\mathnormal{C})}{\partial\mathbf{F}} 
    \quad &&\text{and} \quad
    \mu &&= \frac{\partial\Psi(\mathbf{F},\mathnormal{C})}{\partial\mathnormal{C}} \label{eqn:ConstitutiveRelationBulk}
    \\
    \widetilde{\mathbf{P}} &= \frac{\partial\widetilde{\Psi}(\widetilde{\mathbf{F}},\widetilde{\mathnormal{C}})}{\partial\widetilde{\mathbf{F}}}
    \quad &&\text{and} \quad
    \widetilde{\mu} &&= \frac{\partial\widetilde{\Psi}(\widetilde{\mathbf{F}},\widetilde{\mathnormal{C}})}{\partial\widetilde{\mathnormal{C}}} \label{eqn:ConstitutiveRelationSurf}
\end{alignat}
\end{subequations}
for the bulk and surface first Piola-Kirchhoff stresses tensors, and the bulk and surface chemical potentials, respectively.

For the fifth term in \Cref{eqn:DissipationInequalityChainRule}, the most obvious way to guarantee the dissipation inequality on surface is to impose the condition \citep{mcbride2011geometrically},
\begin{equation}\label{eqn:ChemicalSlavery}
    \mu(\mathbf{X},t) = \widetilde{\mu}(\widetilde{\textbf{X}},t)
    \quad \text{on} \quad \mathnormal{S}
\end{equation}
which prescribes local chemical equilibrium between the surface and the bulk. This point is discussed in \citet{mcbride2011geometrically}, where also the possibility of an alternate strategy to satisfy that constraint is highlighted.

Finally, to maintain that the last two terms in \Cref{eqn:DissipationInequalityChainRule} remain to be negative or zero, we adopt kinetic laws for diffusion \citep{hong2008theory} in bulk and on surface. This allows us to maintain negative semi-definiteness and describes the consistent species diffusion that is driven by gradients of chemical potential:
\begin{subequations}\label{eqn:KineticLaw}
\begin{align}
    \mathbf{J}&=-\mathbf{M}\boldsymbol{\nabla}_{\mathbf{X}}\mu \label{eqn:KineticLawBulk}
    \\
    \widetilde{\mathbf{J}}&=-\widetilde{\mathbf{M}}\widetilde{\boldsymbol{\nabla}}_{\widetilde{\textbf{X}}}\widetilde{\mu} \label{eqn:KineticLawSurf}
\end{align}
\end{subequations}
where $\mathbf{M}$ and $\widetilde{\mathbf{M}}$ are the bulk and surface mobility tensors. In the upcoming section, we will specialize the symmetric and positive definite mobility tensors $\mathbf{M}$ and $\widetilde{\mathbf{M}}$ to fully define the constitutive laws for the fluxes.

\section{Specific considerations for hydrogels}\label{sec:Specific_considerations_for_hydrogels}
In this work, we will focus on a coupled bulk and surface poroelastic framework for hydrogels. We have to specialize our choices for the surface and bulk free energy densities, corresponding constitutive laws, and definition of mobility tensors for our theory to be complete and to be able to proceed to the development of the numerical solution scheme.

\subsection{Particularizing the surface and bulk free energy densities}
For the free energy densities of polymer in the bulk, we adopt the Flory-Huggins model \cite{flory1942thermodynamics,flory1961thermodynamic,treloar1975physics,hong2008theory}, assuming the following additive decomposition into elastic and mixing contributions
\begin{equation}\label{eqn:BulkEnergyDecomp}
      \Psi(\mathbf{F},\mathnormal{C}) = \Psi_{e}(\mathbf{F})+ \Psi_{m}(\mathnormal{C}) \,.
\end{equation}
These can be individually specialized to
\begin{subequations}\label{eqn:StrainEnergyFunctionBulk}
\begin{align}
     \Psi_{e}(\mathbf{F}) &= \frac{\mathnormal{1}}{2}\mathnormal{N}\mathnormal{k}_{B}\mathnormal{T}(\mathnormal{I}_{1}-3-2\ln{\mathnormal{J}}) \label{eqn:StrainEnergyFunctionBulk(a)}\\
     \Psi_{m}(\mathnormal{C}) &= 
     - \frac{\mathnormal{k}_{B}\mathnormal{T}}{\Omega}\left[\Omega\mathnormal{C}\,\ln\left(\frac{1+\Omega\mathnormal{C}}{\Omega\mathnormal{C}}\right)+\frac{\chi}{1+\Omega\mathnormal{C}}\right] \label{eqn:StrainEnergyFunctionBulk(b)}
     \end{align}
\end{subequations}
where $\Omega$ is the volume of a solvent molecule, $\chi$ is a dimensionless parameter of polymer-solvent mixing, $\mathnormal{N}$ is the number of polymer chains per unit reference volume, $\mathnormal{k}_{B}$ is Boltzmann's constant, and $\mathnormal{T}$ is the absolute temperature. 
Note that the first and second terms in \Cref{eqn:StrainEnergyFunctionBulk(b)} represent the entropy and enthalpy of mixing, respectively \cite{hong2008theory}. 

Following \cite{hong2008theory,hong2009inhomogeneous,bouklas2012swelling,bouklas2015nonlinear,ang2020effect}, we assume that the  polymer chains and the diffusion species are individually incompressible. Furthermore, the gel is a condensed matter with negligible void space, so any volume change of the hydrogel is due to species diffusion, thus
\begin{equation}\label{eqn:MolecularIncompressiblity}
    1 + \Omega\mathnormal{C} = \mathnormal{J}
    \Rightarrow
    \mathnormal{C} = \frac{\mathnormal{J} - 1}{\Omega} \,. 
\end{equation}

Similar to \Cref{eqn:BulkEnergyDecomp}, we assume that the surface free energy density of the hydrogel is also decomposed into elastic and mixing contributions as (see \Cref{Rmk:Remark2}))
\begin{equation}\label{eqn:StrainEnergyFunctionSurf}
    \widetilde{\Psi}(\widetilde{\mathbf{F}},\widetilde{\mathnormal{C}}) = \widetilde{\Psi}_{e}(\widetilde{\mathbf{F}},\widetilde{\mathnormal{C}})
    + \widetilde{\beta}\widetilde{\Psi}_{m}(\widetilde{\mathnormal{C}})
\end{equation}
which can be individually expressed as
\begin{subequations}\label{eqn:StrainEnergyFunctionSurf}
\begin{align}
    \widetilde{\Psi}_{e}(\widetilde{\mathbf{F}},\widetilde{\mathnormal{C}}) 
    &= 
    \frac{\widetilde{\kappa}}{2}\left(\widetilde{\mathnormal{J}}-1-\widetilde{\Omega}\widetilde{\mathnormal{C}}\right)^{2}
    + \widetilde{\gamma}\widetilde{\mathnormal{J}} \label{eqn:StrainEnergyFunctionSurf(a)}
    \\
    \widetilde{\Psi}_{m}(\widetilde{\mathnormal{C}}) 
    &= 
    - \frac{\mathnormal{k}_{B}\mathnormal{T}}{\widetilde{\Omega}}\left[\widetilde{\Omega}\widetilde{\mathnormal{C}}\,\ln\left({\frac{1+\widetilde{\Omega}\widetilde{\mathnormal{C}}}{\widetilde{\Omega}\widetilde{\mathnormal{C}}}}\right)+\frac{\widetilde{\chi}}{1+\widetilde{\Omega}\widetilde{\mathnormal{C}}}\right] \label{eqn:StrainEnergyFunctionSurf(b)}
\end{align}
\end{subequations}
In a microscopic description, the length scales of polymer chains and solvent molecules are very different. Focusing on the surface, where the interaction of polymer chains and solvent molecules is crucial, and more specifically the finite thickness of the surface zone, here we consider the the idealization of a constant number of layers $\widetilde{\beta}$ of potential sites that solvent molecules can occupy in the surface zone. Here, $\widetilde{\Omega}$ is the  area that a solvent molecule occupies on the surface, $\widetilde{\chi}$ is a dimensionless parameter of polymer-solvent mixing. In the second term of the right hand side of \Cref{eqn:StrainEnergyFunctionSurf(a)}, we consider a constant surface energy per unit current area $\widetilde{\gamma} \,[\mathrm{J}/\mathrm{m}^{2}]$, which leads to a fluid-like response for the aggregate. Additionally, it is not realistic to enforce the equivalent of exact incompressibility for the surface, as one cannot constrain the number of layers of species on the surface, even though the constituents are individually incompressible. Thus, we introduce a penalty term to loosely introduce a constraint for species concentration in the surface connected with the area change of the boundary, with a penalty coefficient $\widetilde{\kappa}\,[\mathrm{J}/\mathrm{mol}/\mathrm{m}^{2}]$.

\begin{remark}[The application of Flory-Huggins model to surface poroelasticity]\label{Rmk:Remark2}
In the Flory-Huggins model \cite{huggins1941solutions,flory1942thermodynamics}, the solvent and polymer molecules are considered to be arranged in the 3D lattice sites such that each site may be occupied either by a solvent molecule or by a segment of the polymer chain, and the equations of entropy and enthalpy of mixing are derived in the statistical thermodynamics approach (see Chapter 7 of \cite{treloar1975physics} for the details). Without loss of generality, the Flory-Huggins model can be also considered on the 2D lattice model, where the probability of encountering a site occupied by the solvent and polymer molecules is the function of the area. Although the number of adjacent sites would be different in 2D and 3D models, the Flory parameters $\chi$ and $\widetilde{\chi}$ are introduced to  eliminate the number of adjacent sites, and their dependence on the dimensionality is to be fitted by experimental data. We introduce the parameter $\widetilde{\beta}$ in the mixing part of surface free energy $\widetilde{\Psi}_{m}$ to account for the number of 2D lattice layers that arise due to the size difference of solvent molecules and polymer chains considering the surface zone, but we do not consider the interaction energy between the layers. Note that the elastic part of surface free energy $\widetilde{\Psi}_{e}$ does not include the number of layer parameter $\widetilde{\beta}$ which we assume to be a constant.
\end{remark}

\subsection{Constitutive relations}
Using \Cref{eqn:ConstitutiveRelation,eqn:StrainEnergyFunctionBulk,eqn:StrainEnergyFunctionSurf}, the specific constitutive relations are obtained as follows:
\begin{subequations}\label{eqn:SpecificConstitutiveRelation}
\begin{align}
    \mathbf{P} 
    &= 
    \mathnormal{N}\mathnormal{k}_{B}\mathnormal{T}\left(\mathbf{F}+\alpha_{1}\mathnormal{J}\mathbf{F}^{-\mathrm{T}}\right) 
    \quad \text{with} \quad
    \alpha_{1} = -\frac{1}{\mathnormal{J}}+\frac{1}{\mathnormal{N}\Omega}\left[\frac{1}{\mathnormal{J}}+\ln\left(\frac{\mathnormal{J}-1}{\mathnormal{J}}\right)+\frac{\chi}{\mathnormal{J}^{2}}
    \right] 
    \label{eqn:SpecificConstitutiveRelationPK1Bulk}
    \\
    \mu &= 
    \mathnormal{k}_{B}\mathnormal{T}\left[\ln\left(\frac{\Omega\mathnormal{C}}{1+\Omega\mathnormal{C}}\right)+\frac{1}{1+\Omega\mathnormal{C}}+\frac{\chi}{\left(1+\Omega\mathnormal{C}\right)^{2}}\right] \label{eqn:SpecificConstitutiveRelationMuBulk}
    \\
    \widetilde{\mathbf{P}} &=
    \left[
    \widetilde{\kappa}\left(\widetilde{\mathnormal{J}}-1-\widetilde{\Omega}\widetilde{\mathnormal{C}}\right)
    +\widetilde{\gamma}
    \right]
    \widetilde{\mathnormal{J}}\widetilde{\mathbf{F}}^{-\mathrm{T}}
    \label{eqn:SpecificConstitutiveRelationPK1Surf}
    \\
    \widetilde{\mu} &=
    \mathnormal{k}_{B}\mathnormal{T}\left[
    \ln\left(\frac{\widetilde{\Omega}\widetilde{\mathnormal{C}}}{1+\widetilde{\Omega}\widetilde{\mathnormal{C}}}\right)
    +\frac{1}{1+\widetilde{\Omega}\widetilde{\mathnormal{C}}}
    +\frac{\widetilde{\chi}}{\left(1+\widetilde{\Omega}\widetilde{\mathnormal{C}}\right)^{2}}
    \right]
    - \widetilde{\Omega}\,\widetilde{\kappa}\left(\widetilde{J}-1-\widetilde{\Omega}\widetilde{\mathnormal{C}}\right)\
    \Rightarrow 
    \widetilde{\mathnormal{C}} = \widetilde{\mathnormal{C}}\,(\widetilde{\mu},\widetilde{\mathnormal{J}})
    \label{eqn:SpecificConstitutiveRelationMuSurf}
\end{align}
\end{subequations}
which are the specific forms of the constitutive relations for the first Piola-Kirchhoff stresses and the chemical potential in the bulk and on the surface. Note that we do not use the constitutive relation of \Cref{eqn:SpecificConstitutiveRelationMuBulk}, but the incompressibility condition of \Cref{eqn:MolecularIncompressiblity} to determine the bulk concentration $\mathnormal{C}$. On the other hand, we do not have an incompressibility condition for solvent molecules on surface, and we should use the constitutive relation \Cref{eqn:SpecificConstitutiveRelationMuSurf} to determine the surface concentration $\widetilde{\mathnormal{C}}$. However, there is no closed-form solution for the surface concentration in \Cref{eqn:SpecificConstitutiveRelationMuSurf}, so we need to solve the constitutive relation numerically.

\subsection{Bulk and surface diffusion}
Particularizing the expression for the bulk mobility, again following \cite{hong2008theory}, and prescribing an equivalent definition for surface mobility
\begin{subequations}\label{eqn:KineticLaw}
\begin{align}
    \mathbf{M} &= \frac{\mathnormal{C}\mathnormal{D}}{\mathnormal{k}_{B}\mathnormal{T}}\mathbf{C}^{-1} \label{eqn:KineticLawBulk}
    \\
    \widetilde{\mathbf{M}} &= \frac{\widetilde{\mathnormal{C}}\widetilde{\mathnormal{D}}}{\mathnormal{k}_{B}\mathnormal{T}}\widetilde{\mathbf{C}}^{-1} \label{eqn:KineticLawSurf}
\end{align}
\end{subequations}
where $\mathnormal{k}_{B}$ is the Boltzmann’s constant, $\mathnormal{T}$ is an absolute temperature. The coefficient of diffusion of the solvent molecules $\mathnormal{D}$ in the bulk and $\widetilde{\mathnormal{D}}$ on the surface are assumed to be isotropic and independent of the deformation and concentration as the simplest approximation \cite{hong2008theory}. Note that the current concentrations are related to the nominal concentrations as $\mathnormal{c} = \mathnormal{C}/\mathnormal{J}$ and $\widetilde{\mathnormal{c}} = \widetilde{\mathnormal{C}}/\widetilde{\mathnormal{J}}$.

\section{Mixed finite element formulation}\label{sec:Mixed_finite_element_formulation}
This section presents a finite element formulation based on the nonlinear theory in \Cref{sec:A_coupled_continuum_theory,sec:Specific_considerations_for_hydrogels}. The main aim here is to provide an accessible open-source implementation that will be utilized by the research community, and thus, we choose  FEniCS \cite{LoggMardalEtAl2012a,AlnaesBlechta2015a} for the implementation; a choice which also affects the set-up of the mixed finite element formulation. The formulation starts with the strong form of the governing equations and initial and boundary conditions. We first introduce the weak form of the problem and subsequently describe the normalization, discretization, solution steps, and some specific implementation details for FEniCS. As the main focus here is the coupled mechanics between diffusion and deformation in bulk and on surface,  we here-on neglect the source terms, i.e., $\mathnormal{r}=\mathnormal{i}=0$.

\subsection{Three-field weak form}
Since the chemical boundary conditions for hydrogels are often specified in terms of chemical potential or diffusion flux (proportional to the gradient of chemical potential), it is convenient to use the chemical potential (instead of solvent concentration) as an independent variable in the finite element formulation. Additionally, this allows us to avoid $\mathcal{C}^{1}$ continuity requirements \cite{bouklas2015nonlinear}. For this purpose, we rewrite the free energy densities as a function of the deformation gradient and chemical potential through a Legendre transform \cite{hong2008theory,mcbride2011geometrically,bouklas2015nonlinear}, which replaces a variable with its thermodynamic conjugate:
\begin{subequations}\label{eqn:TransformedStrainEnergyFunction}
\begin{align}
    \Phi(\mathbf{F},\mu) &= \Psi(\mathbf{F},\mathnormal{C}) - \mu\mathnormal{C} \\
    \widetilde{\Phi}(\widetilde{\mathbf{F}},\widetilde{\mu}) &= \widetilde{\Psi}(\widetilde{\mathbf{F}},\widetilde{\mathnormal{C}}) - \widetilde{\mu}\widetilde{\mathnormal{C}}
\end{align}
\end{subequations}
After the Legendre transform, similar to \Cref{eqn:ConstitutiveRelation}, the constitutive relations can be rewritten as follows:
\begin{subequations}\label{eqn:TransformedConstitutiveRelation}
\begin{alignat}{3}
    \mathbf{P} &= \frac{\partial\Phi(\mathbf{F},\mu)}{\partial\mathbf{F}} 
    \quad &&\text{and} \quad
    \mathnormal{C} &&= -\frac{\partial\Phi(\mathbf{F},\mu)}{\partial\mu} \label{eqn:ConstitutiveRelationBulk}
    \\
    \widetilde{\mathbf{P}} &= \frac{\partial\widetilde{\Phi}(\widetilde{\mathbf{F}},\widetilde{\mu})}{\partial\widetilde{\mathbf{F}}}
    \quad &&\text{and} \quad
    \widetilde{\mathnormal{C}} &&= -\frac{\partial\widetilde{\Phi}(\widetilde{\mathbf{F}},\widetilde{\mu})}{\partial\widetilde{\mu}} \label{eqn:ConstitutiveRelationSurf}
\end{alignat}
\end{subequations}
which yields the same constitutive relation for the surface first Piola-Kirchhoff $\widetilde{\mathbf{P}}$, but the bulk first Piola-Kirchhoff stress $\mathbf{P}$ should be modified because the incompressibility condition was enforced by substituting the \Cref{eqn:MolecularIncompressiblity} into \Cref{eqn:StrainEnergyFunctionBulk} to eliminate the bulk concentration $\mathnormal{C}$ \cite{hong2009inhomogeneous}:
\begin{align}
    \mathbf{P} &=
    \mathnormal{N}\mathnormal{k}_{B}\mathnormal{T}\left(\mathbf{F}+\alpha_{2}\mathnormal{J}\mathbf{F}^{-\mathrm{T}}\right) 
    \quad \text{with} \quad \alpha_{2} = -\frac{1}{\mathnormal{J}}+\frac{1}{\mathnormal{N}\Omega}\left[\frac{1}{\mathnormal{J}}+\ln\left(\frac{\mathnormal{J}-1}{\mathnormal{J}}\right)+\frac{\chi}{\mathnormal{J}^{2}}-\frac{\mu}{\mathnormal{k}_{B}\mathnormal{T}}\right]
    \label{eqn:TransformedConstitutiveRelationBulk}
\end{align}
Note that the bulk concentration $\mathnormal{C}$ can still be obtained by incompressibility condition of \Cref{eqn:MolecularIncompressiblity}, but the surface concentration $\widetilde{\mathnormal{C}}$ is now given implicitly by solving a nonlinear algebraic equation in \Cref{eqn:SpecificConstitutiveRelationMuSurf}.

Keeping into account the inability of the FEniCS framework to solve nonlinear equations at the Gauss-point level, we proceed to solve \Cref{eqn:SpecificConstitutiveRelationMuSurf} directly with the mixed finite element method. The weak form of the problem is obtained by using a set of test functions, which satisfy the necessary integrability conditions. By multiplying \Cref{eqn:EquilibriumBulk}, \Cref{eqn:ConservationLawBulk} and \Cref{eqn:SpecificConstitutiveRelationMuSurf} with the test functions $\delta\mathbf{u}$, $\delta\mu$ and $\delta\widetilde{\mathnormal{C}}$, and integrating over the domain, respectively, then we obtain that
\begin{subequations}\label{eqn:WeakForm}
\begin{align}
    &\int_{\mathnormal{V}} \mathbf{P}:\boldsymbol{\nabla}_{\mathbf{X}}\delta\mathbf{u} \, \mathrm{d}V
    +\int_{\mathnormal{S}} \widetilde{\mathbf{P}}:\widetilde{\boldsymbol{\nabla}}_{\widetilde{\textbf{X}}}\delta\mathbf{u} \, \mathrm{d}S
    =0 \label{eqn:WeakForm_Equilibrium} \\
    &\int_{\mathnormal{V}} \dot{\mathnormal{C}}\,\delta\mu \, \mathrm{d}V -\int_{\mathnormal{V}} \mathbf{J}\cdot\boldsymbol{\nabla}_{\mathbf{X}}\delta\mu \, \mathrm{d}V 
    +\int_{\mathnormal{S}} \dot{\widetilde{\mathnormal{C}}}\,\delta\mu \, \mathrm{d}S 
    - \int_{\mathnormal{S}} \widetilde{\mathbf{J}}\cdot\widetilde{\boldsymbol{\nabla}}_{\widetilde{\textbf{X}}}\delta\mu\, \mathrm{d}S 
    = 0 \label{eqn:WeakForm_ConservationLaw} \\
    &\int_{\mathnormal{S}} \Bigg[ \widetilde{\mu}-\mathnormal{k}_{B}\mathnormal{T}\Big[
    \ln\left(\frac{\widetilde{\Omega}\widetilde{\mathnormal{C}}}{1+\widetilde{\Omega}\widetilde{\mathnormal{C}}}\right)
    +\frac{1}{1+\widetilde{\Omega}\widetilde{\mathnormal{C}}}
    +\frac{\widetilde{\chi}}{\left(1+\widetilde{\Omega}\widetilde{\mathnormal{C}}\right)^{2}}
    \Big]
    - \widetilde{\Omega}\,\widetilde{\kappa}\left(\widetilde{J}-1-\widetilde{\Omega}\widetilde{\mathnormal{C}}\right)\Bigg]\delta\widetilde{\mathnormal{C}} \, \mathrm{d}S 
    = 0 \label{eqn:WeakForm_ConstitutiveRelation} 
\end{align}
\end{subequations}
where the derivation of \Cref{eqn:WeakForm_ConservationLaw} can be found in \Cref{sec:Derivation of weak form of species conservation}. The statement of the weak form is to find the trial functions, $\mathbf{u}$, $\mu$ and $\widetilde{\mathnormal{C}}$, such that the integrals in \Cref{eqn:WeakForm} are satisfied for any permissible test functions, $\delta\mathbf{u}$, $\delta\mu$ and $\delta\widetilde{\mathnormal{C}}$.

\subsection{Normalization}\label{sec:Normalization}
For the finite element simulations in the following section, all variables and parameters are normalized, as denoted by $\{\widehat{\bullet}\}$. All lengths are normalized by a characteristic dimension, $\mathnormal{H}$ (e.g. the length of the edge of a cube in the reference configuration). The chemical potential and stresses are normalized as follows,
\begin{equation}
    \widehat{\mu} = \frac{\mu}{\mathnormal{k}_{B}\mathnormal{T}}
    ,\quad
    \widehat{\mathbf{P}} = \frac{\mathbf{P}}{\mathnormal{N}\mathnormal{k}_{B}\mathnormal{T}}
    ,\quad
    \widehat{\widetilde{\mathbf{P}}} = \frac{\widetilde{\mathbf{P}}}{\mathnormal{N}\mathnormal{k}_{B}\mathnormal{T}\mathnormal{H}}
\end{equation}
Following the normalization of the surface stress, surface energy and penalty coefficient should be normalized in the same way,
\begin{equation}
    \widehat{\widetilde{\gamma}} = \frac{\widetilde{\gamma}}{\mathnormal{N}\mathnormal{k}_{B}\mathnormal{T}\mathnormal{H}}
    ,\quad
    \widehat{\widetilde{\kappa}} = \frac{\widetilde{\kappa}}{\mathnormal{N}\mathnormal{k}_{B}\mathnormal{T}\mathnormal{H}}
\end{equation}
Solvent concentration and time are normalized as follows,
\begin{equation}
    \widehat{\mathnormal{C}} = \Omega\mathnormal{C}
    ,\quad
    \widehat{\widetilde{\mathnormal{C}}} = \widetilde{\Omega}\widetilde{\mathnormal{C}}
    ,\quad
    \hat{\mathnormal{t}} = \frac{\mathnormal{t}}{\tau}
\end{equation}
where $\tau=\mathnormal{H}^{2}/\mathnormal{D}$ is the characteristic time scale of diffusion. 

Recall that the  shear modulus linearized about the undeformed dry state ($\mathbf{F}=\mathbf{I}$) is $\mathnormal{G}=\mathnormal{N}\mathnormal{k}_{B}\mathnormal{T}$ \cite{bouklas2012swelling}. In a homogeneously swollen stress-free state, the deformation gradient takes the form $\mathbf{F}=\lambda_0\mathbf{I}$, with $\lambda_0$ the swelling stretch, the  shear modulus about that state is defined as follows \cite{bouklas2012swelling}:
\begin{equation}
    \mathnormal{G}_{0}=\frac{\mathnormal{N}\mathnormal{k}_{B}\mathnormal{T}}{\lambda_{0}}
\end{equation}
Considering again a homogeneously swollen state, the elastocapillary length scale is defined by $\ell = \widetilde{\gamma}/\mathnormal{G}_{0}$ for constant surface free energy per unit current surface area. Taking into account the characteristic dimension $\mathnormal{H}$ the normalized elastocapillary length scale is given as
\begin{equation}
    \widehat{\ell} = \frac{\ell}{\mathnormal{H}} = \widehat{\widetilde{\gamma}}\lambda_{0}
\end{equation}
By substituting the normalized variables into the weak forms in \Cref{eqn:WeakForm}, we can obtain the normalized weak forms, for the bulk and the surface.
\begin{subequations}\label{eqn:NormalizedWeakForm} 
\begin{align}
    &\int_{\mathnormal{V}}\widehat{\mathbf{P}}:\boldsymbol{\nabla}_{\widehat{\textbf{X}}}\delta\widehat{\mathbf{u}}\,\mathrm{d}\widehat{V} +\int_{\mathnormal{S}} \widehat{\widetilde{\mathbf{P}}}:\widetilde{\boldsymbol{\nabla}}_{\widehat{\widetilde{\textbf{X}}}}\delta\widehat{\mathbf{u}} \, \mathrm{d}\widehat{S} = 0 \label{eqn:NormalizedWeakForm_Equilibrium}
    \\
    &\frac{\widetilde{\Omega}\mathnormal{H}}{\Omega}\int_{\mathnormal{V}} \left(\frac{\mathrm{d}\widehat{\mathnormal{C}}}{\mathrm{d}\hat{\mathnormal{t}}}\,\delta\widehat{\mu}
    - \int_{\mathnormal{V}} \widehat{\mathbf{J}}\cdot\boldsymbol{\nabla}_{\widehat{\mathbf{X}}}\delta\widehat{\mu} \right)\mathrm{d}\widehat{V}
    + \int_{\mathnormal{S}} \left(\frac{\mathrm{d}\widehat{\widetilde{\mathnormal{C}}}}{\mathrm{d}\hat{\mathnormal{t}}}\,\delta\widehat{\widetilde{\mu}}
    - \int_{\mathnormal{S}} \widehat{\widetilde{\mathbf{J}}}\cdot\widetilde{\boldsymbol{\nabla}}_{\widehat{\widetilde{\textbf{X}}}}\delta\widehat{\widetilde{\mu}} \right)\mathrm{d}\widehat{S} = 0  \label{eqn:NormalizedWeakForm_Conservation}
    \\
    &\int_{\mathnormal{S}} \Bigg[\widehat{\widetilde{\mu}}-\ln\left(\frac{\widehat{\widetilde{\mathnormal{C}}}}{1+\widehat{\widetilde{\mathnormal{C}}}}\right)
    -\frac{1}{1+\widehat{\widetilde{\mathnormal{C}}}}
    -\frac{\widetilde{\chi}}{\left(1+\widehat{\widetilde{\mathnormal{C}}}\right)^{2}}
    +\mathnormal{N}\widetilde{\Omega}\mathnormal{H}\,\widehat{\widetilde{\kappa}}\left(\widetilde{J}-1-\widehat{\widetilde{\mathnormal{C}}}\right)\Bigg]\delta\widehat{\widetilde{\mathnormal{C}}}\, \mathrm{d}\widehat{S} = 0 \label{eqn:NormalizedWeakForm_ConcentrationSurf} 
\end{align}
\end{subequations}
where $\mathnormal{N}\Omega$ and $\mathnormal{N}\widetilde{\Omega}\mathnormal{H}$ are the dimensionless parameters, and the normalized fluxes in bulk and on surface are obtained by
\begin{subequations}\label{eqn:NormalizedFluxes}
\begin{align}
    \widehat{\mathbf{J}}
    &= -\widehat{\mathnormal{C}}\mathbf{C}^{-1}\cdot\boldsymbol{\nabla}_{\widehat{\mathbf{X}}}\widehat{\mu}
    \\
    \widehat{\widetilde{\mathbf{J}}}
    &= -\widehat{\widetilde{\mathnormal{C}}}\frac{\widetilde{\mathnormal{D}}}{\mathnormal{D}}\widetilde{\mathbf{C}}^{-1}\cdot\widetilde{\boldsymbol{\nabla}}_{\widehat{\widetilde{\textbf{X}}}}\widehat{\widetilde{\mu}}
\end{align}
\end{subequations}
where $\widetilde{\mathnormal{D}}/\mathnormal{D}$ is the ratio of diffusivity between in bulk and on surface.

\subsection{Temporal discretization}
The backward Euler scheme is used to integrate \Cref{eqn:NormalizedWeakForm_Conservation} over time:
\begin{align}
    &\frac{\widetilde{\Omega}\mathnormal{H}}{\Omega}\int_{\mathnormal{V}} \left[\frac{1}{\Delta\hat{\mathnormal{t}}}\left(\widehat{\mathnormal{C}}{}^{\hat{\mathnormal{t}}+\Delta\hat{\mathnormal{t}}}-\widehat{\mathnormal{C}}{}^{\hat{\mathnormal{t}}}\right)\delta\widehat{\mu}
    - \int_{\mathnormal{V}} \widehat{\mathbf{J}}{}^{\hat{\mathnormal{t}}+\Delta\hat{\mathnormal{t}}}\cdot\boldsymbol{\nabla}_{\widehat{\mathbf{X}}}\delta\widehat{\mu} \right] \mathrm{d}\widehat{V}
    \nonumber\\
    &+ \int_{\mathnormal{S}} \left[ \frac{1}{\Delta\hat{\mathnormal{t}}}\left(\widehat{\widetilde{\mathnormal{C}}}{}^{\hat{\mathnormal{t}}+\Delta\hat{\mathnormal{t}}}-\widehat{\widetilde{\mathnormal{C}}}{}^{\hat{\mathnormal{t}}}\right)\delta\widehat{\widetilde{\mu}}
    - \int_{\mathnormal{S}} \widehat{\widetilde{\mathbf{J}}}{}^{\hat{\mathnormal{t}}+\Delta\hat{\mathnormal{t}}}\cdot\widetilde{\boldsymbol{\nabla}}_{\widehat{\widetilde{\textbf{X}}}}\delta\widehat{\widetilde{\mu}}\right] \mathrm{d}\widehat{S}
    = 0 \label{eqn:NormalizedWeakForm_BackwardEuler}
\end{align}
where the superscripts indicate the time step, at the current time step ($\hat{\mathnormal{t}}+\Delta\hat{\mathnormal{t}}$) or the previous step $\hat{\mathnormal{t}}$. We can combine \Cref{eqn:NormalizedWeakForm_Equilibrium}, \Cref{eqn:NormalizedWeakForm_ConcentrationSurf} and \Cref{eqn:NormalizedWeakForm_BackwardEuler} as
\begin{align}\label{eqn:CombinedWeakForm}
    &\int_{\mathnormal{V}} \widehat{\mathbf{P}}:\boldsymbol{\nabla}_{\widehat{\mathbf{X}}}\delta\widehat{\mathbf{u}} \, \mathrm{d}\widehat{V} + \int_{\mathnormal{S}} \widehat{\widetilde{\mathbf{P}}}:\widetilde{\boldsymbol{\nabla}}_{\widehat{\widetilde{\textbf{X}}}}\delta\widehat{\mathbf{u}} \, \mathrm{d}\widehat{S}  
    \nonumber\\
    &+ \frac{\widetilde{\Omega}\mathnormal{H}}{\Omega}\int_{\mathnormal{V}} \left[\left(\widehat{\mathnormal{C}} - \widehat{\mathnormal{C}}{}^{\hat{\mathnormal{t}}}\right)\delta\widehat{\mu} - \Delta\hat{\mathnormal{t}}\,\widehat{\mathbf{J}}\cdot\boldsymbol{\nabla}_{\widehat{\mathbf{X}}}\delta\widehat{\mu}\right]\, \mathrm{d}\widehat{V}
    + \int_{\mathnormal{S}} \left[\left(\widehat{\widetilde{\mathnormal{C}}} - \widehat{\widetilde{\mathnormal{C}}}{}^{\hat{\mathnormal{t}}}\right)\delta\widehat{\widetilde{\mu}} - \Delta\hat{\mathnormal{t}}\,\widehat{\widetilde{\mathbf{J}}}\cdot\widetilde{\boldsymbol{\nabla}}_{\widehat{\widetilde{\textbf{X}}}}\delta\widehat{\widetilde{\mu}}\right]\, \mathrm{d}\widehat{S} 
    \nonumber\\
    &+\int_{\mathnormal{S}} \Bigg[\widehat{\widetilde{\mu}}-\ln\left(\frac{\widehat{\widetilde{\mathnormal{C}}}}{1+\widehat{\widetilde{\mathnormal{C}}}}\right)
    -\frac{1}{1+\widehat{\widetilde{\mathnormal{C}}}}
    -\frac{\widetilde{\chi}}{\left(1+\widehat{\widetilde{\mathnormal{C}}}\right)^{2}}
    +\mathnormal{N}\widetilde{\Omega}\mathnormal{H}\,\widehat{\widetilde{\kappa}}\left(\widetilde{J}-1-\widehat{\widetilde{\mathnormal{C}}}\right)\Bigg]\delta\widehat{\widetilde{\mathnormal{C}}}\, \mathrm{d}\widehat{S}
    = 0
\end{align}
where the superscript ($\hat{\mathnormal{t}}+\Delta\hat{\mathnormal{t}}$) is omitted for all the terms at the current time step and $\widehat{\mathnormal{C}}{}^{\hat{\mathnormal{t}}}$ and $\widehat{\widetilde{\mathnormal{C}}}{}^{\hat{\mathnormal{t}}}$ are the species concentration at the previous time step in the bulk and on the surface.

\subsection{Spatial discretization}
A mixed finite element method is utilized to solve for the normalized displacement, chemical potential and surface concentration fields concurrently. To avoid the numerical instability with the mixed method, we should employ proper spatial discretization techniques \cite{bouklas2015nonlinear}. We utilize a Taylor-Hood element \cite{taylor1973numerical} for the bulk quantities (displacement and chemical potential), where the interpolation order for chemical potential (linear) is one-order lower than for the displacement (quadratic). We note that a linear interpolation is utilized for the and surface concentration field.

The normalized displacement and chemical potential are interpolated through the domain of interest as
\begin{equation}\label{eqn:SpatialDiscretization}
    \widehat{\mathbf{u}} = \mathbf{H}{}^{\widehat{\mathbf{u}}}\widehat{\mathbf{u}}{}^{n},
    \quad
    \widehat{\mu} = \mathbf{H}{}^{\widehat{\mu}}\widehat{\boldsymbol{\mu}}{}^{n},
    \quad
    \widehat{\widetilde{\mathnormal{C}}} = \widetilde{\mathbf{H}}{}^{\widehat{\widetilde{\mathnormal{C}}}}\widehat{\widetilde{\boldsymbol{\mathnormal{C}}}}{}^{n}
\end{equation}
where $\mathbf{H}{}^{\widehat{\mathbf{u}}}$, $\mathbf{H}{}^{\widehat{\mu}}$ and $\widetilde{\mathbf{H}}{}^{\widehat{\widetilde{\mathnormal{C}}}}$ are the shape functions, $\widehat{\mathbf{u}}{}^{n}$, $\widehat{\boldsymbol{\mu}}{}^{n}$ and $\widehat{\widetilde{\boldsymbol{\mathnormal{C}}}}{}^{n}$ are the nodal values of the normalized displacement, chemical potential and surface concentration, respectively. Note that the shape function $\widetilde{\mathbf{H}}{}^{\widehat{\widetilde{\mathnormal{C}}}}$ is only defined on surface elements. The test functions are discretized in the same way
\begin{equation}
    \delta\widehat{\mathbf{u}} = \mathbf{H}{}^{\widehat{\mathbf{u}}}\delta\widehat{\mathbf{u}}{}^{n},
    \quad
    \delta\widehat{\mu} = \mathbf{H}{}^{\widehat{\mu}}\delta\widehat{\boldsymbol{\mu}}{}^{n},
    \quad
    \delta\widehat{\widetilde{\mathnormal{C}}} = \widetilde{\mathbf{H}}{}^{\widehat{\widetilde{\mathnormal{C}}}}\delta\widehat{\widetilde{\boldsymbol{\mathnormal{C}}}}{}^{n}
\end{equation}
The stresses, concentrations, and fluxes are evaluated at integration points, depending on the gradients of the displacement and chemical potential via the constitutive relations. Taking the gradient of \Cref{eqn:SpatialDiscretization}, we obtain that
\begin{subequations}
\begin{alignat}{2}
    \boldsymbol{\nabla}_{\widehat{\mathbf{X}}}\widehat{\mathbf{u}} &= \boldsymbol{\nabla}_{\widehat{\mathbf{X}}}\mathbf{H}{}^{\widehat{\mathbf{u}}}\widehat{\mathbf{u}}{}^{n} &&= \mathbf{B}{}^{\widehat{\mathbf{u}}}\widehat{\mathbf{u}}{}^{n} \\
    \boldsymbol{\nabla}_{\widehat{\mathbf{X}}}\widehat{\mu} &= \boldsymbol{\nabla}_{\widehat{\mathbf{X}}}\mathbf{H}{}^{\widehat{\mu}}\delta\widehat{\boldsymbol{\mu}}{}^{n} &&= \mathbf{B}{}^{\widehat{\mu}}\widehat{\boldsymbol{\mu}}{}^{n} \\
    \boldsymbol{\nabla}_{\widehat{\mathbf{X}}}\widehat{\widetilde{\mathnormal{C}}} &= \boldsymbol{\nabla}_{\widehat{\mathbf{X}}}\widetilde{\mathbf{H}}{}^{\widehat{\widetilde{\mathnormal{C}}}}\widehat{\widetilde{\boldsymbol{\mathnormal{C}}}}{}^{n} &&= \widetilde{\mathbf{B}}{}^{\widehat{\widetilde{\mathnormal{C}}}}\widehat{\widetilde{\boldsymbol{\mathnormal{C}}}}{}^{n}
\end{alignat}
\end{subequations}
where $\mathbf{B}{}^{\widehat{\mathbf{u}}}$ and $\mathbf{B}{}^{\widehat{\mu}}$ are the gradients of the shape functions in the bulk, and $\widetilde{\mathbf{B}}{}^{\widehat{\widetilde{\mathnormal{C}}}}$ is the one on the surface.

\subsection{Nonlinear solution}

The weak form in \Cref{eqn:CombinedWeakForm} can be expressed as a system of nonlinear equations,
\begin{equation}\label{eqn:NonlinearEquation}
    \mathcal{N}(\mathbf{d}) = \mathbf{f} \quad \text{with} \quad \mathbf{d}
    = \left[ \widehat{\mathbf{u}}{}^{n} 
    \; 
    \widehat{\boldsymbol{\mu}}{}^{n}
    \;  
    \widehat{\widetilde{\boldsymbol{\mathnormal{C}}}}{}^{n}
    \right]^{\mathrm{T}}
\end{equation}
Note that $\mathcal{N}(\mathbf{d})$ denotes the part of the weak form that is not known at current time step, and we take all the known quantities to the right-hand side and denote as $\mathbf{f}$ (which is known from previous time step). The residual of nonlinear equations in iteration step $\mathnormal{i}$ is given by $\mathbf{R}_{\mathnormal{i}}=\mathbf{f}-\mathcal{N}(\mathbf{d}_{\mathnormal{i}})$, which can be solved using the Newton–Raphson method. In particular, the procedure requires the calculation of the tangent Jacobian matrix at each iteration, namely,
\begin{equation}\label{eqn:BlockJacobianMatrix}
    \frac{\partial\mathcal{N}}{\partial\mathbf{d}} \Bigg\vert_{\mathbf{d}_{i}} 
    = \begin{bmatrix}
    \mathbf{K}^{\widehat{\mathbf{u}}\widehat{\mathbf{u}}} 
    & \mathbf{K}^{\widehat{\mathbf{u}}\widehat{\mu}}
    & \mathbf{K}^{\widehat{\mathbf{u}}\widehat{\widetilde{\mathnormal{C}}}}  
    \\
    \mathbf{K}^{\widehat{\mu}\widehat{\mathbf{u}}} 
    & \mathbf{K}^{\widehat{\mu}\widehat{\mu}}
    & \mathbf{K}^{\widehat{\mu}\widehat{\widetilde{\mathnormal{C}}}} 
    \\
    \mathbf{K}^{\widehat{\widetilde{\mathnormal{C}}}\widehat{\mathbf{u}}} 
    & \mathbf{K}^{\widehat{\widetilde{\mathnormal{C}}}\widehat{\mu}} 
    & \mathbf{K}^{\widehat{\widetilde{\mathnormal{C}}}\widehat{\widetilde{\mathnormal{C}}}} 
    \end{bmatrix}
\end{equation}

FEniCS version 2019.2.0 \cite{LoggMardalEtAl2012a,AlnaesBlechta2015a} is used to numerically solve the coupled non-linear equations via the Portable Extensible Toolkit for Scientific Computations (PETSc) Scalable Nonlinear Equations Solvers (SNES) interface \cite{balay2019petsc}. This process repeats until a level of convergence specified within the SNES solver. At each iteration, the block Jacobian matrices are set up using multiphenics \cite{ballarin2019multiphenics}, a python library that also facilitates the definition of boundary restricted variables within FEniCS, a feature necessary for dealing in our case with the surface concentration. We note that the surface chemical potential is tied with the definition of the corresponding bulk quantity, but the surface concentration is not.

\subsection{FEniCS implementation details}
Although many documents \cite{logg2012automated,langtangen2017solving,bleyer2018numerical} provide FEniCS implementation details, it is still not trivial to deal with surface kinematics using FEniCS as several surface quantities have a rank deficiency. For educational purposes, we review a few important mathematical definitions of surface quantities, and then we provide their corresponding FEniCS definitions.

The surface unit tensor is defined by $\widetilde{\textbf{I}}=\textbf{I}-\textbf{N}\otimes\textbf{N}$, which can be interpreted as surface (idempotent) projection tensor, and surface deformation gradient can be obtained from surface projection of bulk deformation gradient $\widetilde{\textbf{F}}=\mathbf{F}\widetilde{\mathbf{I}}$. Similarly, the surface gradient is obtained by surface projection of the bulk gradient $\widetilde{\boldsymbol{\nabla}}_{\widetilde{\textbf{X}}}\mu=\boldsymbol{\nabla}_{\textbf{X}}\mu\widetilde{\textbf{I}}$. 

\begin{lstlisting}[language=Python]
N = FacetNormal(mesh)
I_bulk = Identity(3)
I_surf = I_bulk - outer(N,N)
F_surf = dot(F_bulk,I_surf)
grad_mu_surf = dot(grad(mu),I_surf)
\end{lstlisting}

Note that the surface unit tensor $\widetilde{\mathbf{I}}$ is absent from the (principal) normal component $\textbf{N}\otimes\textbf{N}$; as a result, the surface projection always leads to a rank deficiency, and we cannot perform the inverse operations for the surface quantities. Nevertheless, they possess an inverse in the generalized sense, i.e., $\widetilde{\textbf{F}}^{-1}:=\widetilde{\mathbf{I}}\mathbf{F}^{-1}$ and $\widetilde{\textbf{C}}^{-1}:=\widetilde{\mathbf{I}}\mathbf{C}^{-1}\widetilde{\mathbf{I}}$.

\begin{lstlisting}[language=Python]
F_surf_inv = dot(I_surf,inv(F_bulk))
C_surf_inv = dot(I_surf,dot(C_bulk,I_surf))
\end{lstlisting}

In addition, we cannot perform the determinant operations for the surface quantities due to rank deficiency. We have the alternative way of using Nanson's formula to obtain the determinant of surface quantities, i.e., $\widetilde{\mathnormal{J}}=\text{det}\,\widetilde{\textbf{F}}:=\lvert\mathrm{cof}\mathbf{F}\mathbf{N}\rvert$. 

\begin{lstlisting}[language=Python]
A_surf = dot(cofac(F_bulk),N)
J_surf = sqrt(dot(A_surf,A_surf))
\end{lstlisting}

\section{Numerical examples}\label{sec:Numerical_examples}
To study the performance of the suggested framework, we analyze the transient responses for two initial boundary value problems involving bulk and surface poroelasticity of hydrogels: (1) free contraction of a cube with smooth edges and (2) uniaxial tension of a cube with sharp edges. In both cases the edge length of the cubes is $10^{-3}\,\mathrm{m}$. In the following numerical study, we take the initial swelling ratio at $\lambda_{0}=3.215$, and the Flory interaction parameter at $\chi=\widetilde{\chi}=0.2$. At room temperature $\mathnormal{k}_{B}\mathnormal{T}=4\times10^{-21}\,\mathrm{J}/\mathrm{mol}$, and the representative value of the volume and area per molecule are $\Omega=10^{-28}\,\mathrm{m}^{3}$ and $\widetilde{\Omega}=10^{-19}\,\mathrm{m}^{2}$. In the absence of solvent molecules, the dry network has shear moduli of $\mathnormal{N}\mathnormal{k}_{B}\mathnormal{T}=4\times10^{4}\,\mathrm{N}/\mathrm{m}^{2}$, which gives $\mathnormal{N}\Omega=10^{-3}$ and $\mathnormal{N}\widetilde{\Omega}\mathnormal{H}=10^{3}$ \cite{ang2020effect}. The characteristic dimension and time scales are set as $\mathnormal{H} = 10^{-3}\,\mathrm{m}$ and $\tau=1.0\,\mathrm{sec}$. Without loss of generality, we prescribe the following values for the parameters $\widetilde{\beta}=1.0$, $\widehat{\widetilde{\gamma}}=1.0$, $\widehat{\widetilde{\kappa}}=10^{-3}$ and $\widetilde{\mathnormal{D}}/\mathnormal{D}=1.0$. 

For robustness of the numerical procedure, we follow a two-stage process. We initialize from a homogeneously swollen state, with initial homogeneous swelling stretch $\lambda_{0}$, neglecting elastic surface energy contributions. In the first stage, we ramp the surface energy $\widehat{\widetilde{\gamma}}$ linearly, from zero to its prescribed value, for the time interval $\mathnormal{t}/\tau \in [0.0,1.0]$. This time interval is  significantly smaller compared to the time the system needs for equilibration. Then in the second stage, we exponentially increase the time steps $\Delta\mathnormal{t}/\tau$ until equilibrium is attained, while holding the value of the surface energy $\widehat{\widetilde{\gamma}}$ fixed.

We denote the free-swelling stretch by $\mathbf{F}=\lambda_{0}\mathbf{I}$, which relates to the chemical potential $\widehat{\mu}_{0}$ by setting stress \Cref{eqn:TransformedConstitutiveRelationBulk} to be zero \cite{hong2008theory,hong2009inhomogeneous},
\begin{equation}\label{eqn:Initial_condition}
    \mathnormal{N}\Omega\left(\frac{1}{\lambda_{0}}-\frac{1}{\lambda_{0}^{3}}\right) + \ln\left(1-\frac{1}{\lambda_{0}^{3}}\right) + \frac{1}{\lambda_{0}^{3}} + \frac{\chi}{\lambda_{0}^{6}} = \widehat{\mu}_{0}
\end{equation}
We can obtain the initial free-swelling stretch $\lambda_{0}=3.215$ by setting the immersed condition, $\widehat{\mu}_{0}=0.0$ \cite{hong2008theory}. We can also obtain the initial normalized surface concentration by numerically solving the normalized constitutive relation of \Cref{eqn:SpecificConstitutiveRelationMuSurf}, which yields the initial value, $\widehat{\widetilde{\mathnormal{C}}}_{0}\approx9.3673$. 
It is important to note that the boundary of swollen hydrogel is assumed to be impermeable for the cases studies here, which is formulated by the flux boundary condition, $\widetilde{\mathbf{J}}_{p}=0$ (see \Cref{eqn:ConservationLaw}). That is, the species on the surface are not allowed to migrate into the exterior (environment). We note that for immersed conditions, we would need to assume local equilibrium instantaneously on the surface which would override the surface diffusion mechanism. 

\begin{figure*}[ht]
    \centering
    \includegraphics[width=0.7\linewidth]{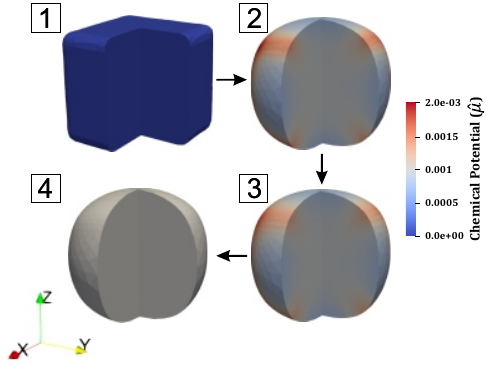}
    \caption{Temporal sequence of the chemical potential during the free contraction of cubes.}
    \label{fig:Fig2}
\end{figure*}

\begin{figure*}[ht]
    \centering
    \includegraphics[width=0.7\linewidth]{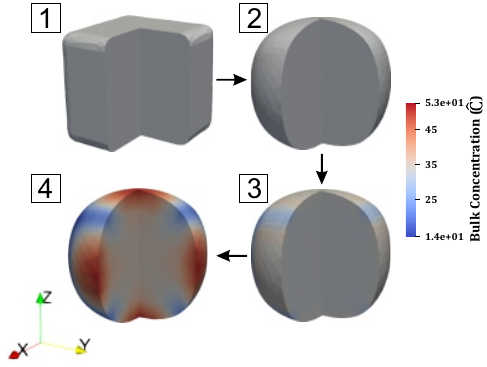}
    \caption{Temporal sequence of the bulk contraction during the free contraction of cubes.
    .}
    \label{fig:Fig3}
\end{figure*}

To investigate the numerical stability when time step is small ($\Delta\mathnormal{t}\rightarrow0$), the block Jacobian matrix in \Cref{eqn:BlockJacobianMatrix} can be reduced to
\begin{equation}\label{eqn:ReducedBlockJacobianMatrix}
    \frac{\partial\mathcal{N}}{\partial\mathbf{d}} \Bigg\vert_{\mathbf{d}_{i}} 
    = \begin{bmatrix}
    \mathbf{K}^{\widehat{\mathbf{u}}\widehat{\mathbf{u}}} 
    & \mathbf{K}^{\widehat{\mathbf{u}}\widehat{\mu}}
    & \mathbf{K}^{\widehat{\mathbf{u}}\widehat{\widetilde{\mathnormal{C}}}} \\
    \mathbf{K}^{\widehat{\mu}\widehat{\mathbf{u}}} 
    & \mathbf{K}^{\widehat{\mu}\widehat{\mu}}
    & \mathbf{K}^{\widehat{\mu}\widehat{\widetilde{\mathnormal{C}}}}  \\
    \mathbf{K}^{\widehat{\widetilde{\mathnormal{C}}}\widehat{\mathbf{u}}}
    & \mathbf{K}^{\widehat{\widetilde{\mathnormal{C}}}\widehat{\mu}}
    & \mathbf{K}^{\widehat{\widetilde{\mathnormal{C}}}\widehat{\widetilde{\mathnormal{C}}}}
    \end{bmatrix}
    \approx \begin{bmatrix}
    \mathbf{K}^{\widehat{\mathbf{u}}\widehat{\mathbf{u}}} 
    & \mathbf{K}^{\widehat{\mathbf{u}}\widehat{\mu}}
    & \mathbf{K}^{\widehat{\mathbf{u}}\widehat{\widetilde{\mathnormal{C}}}} \\
    \mathbf{K}^{\widehat{\mu}\widehat{\mathbf{u}}}
    & \mathbf{0}
    & \mathbf{K}^{\widehat{\mu}\widehat{\widetilde{\mathnormal{C}}}}  \\
    \mathbf{K}^{\widehat{\widetilde{\mathnormal{C}}}\widehat{\mathbf{u}}}
    & \mathbf{K}^{\widehat{\widetilde{\mathnormal{C}}}\widehat{\mu}}
    & \mathbf{K}^{\widehat{\widetilde{\mathnormal{C}}}\widehat{\widetilde{\mathnormal{C}}}}
    \end{bmatrix}
\end{equation}
where $\mathbf{K}^{\widehat{\mu}\widehat{\mu}}$ is proportional to time step ($\Delta\mathnormal{t}$), and it approaches zero at the short time limit. This indicates the saddle point problem structure inherently, which is known to lead to numerical oscillations; however, as the time progresses in the transient process, the parabolic nature of the diffusion equations regularizes the problem (see \cite{wan2003stabilized} Chapter 3 for the details of the saddle point problem of $3\times3$ block matrices). It is worth noting, that this type of oscillations for the bulk poroelasticity problem, stemming from the inf-sup problem are known to be aleviated by several approaches, including the choice of Taylor-Hood spaces (as shown for the nonlinear problem in \cite{bouklas2015nonlinear}). Here we have also chosen a Taylor-Hood space between the displacement and chemical potential, but also between the displacement and surface concentration.

\begin{figure*}[ht]
    \centering
    \includegraphics[width=0.7\linewidth]{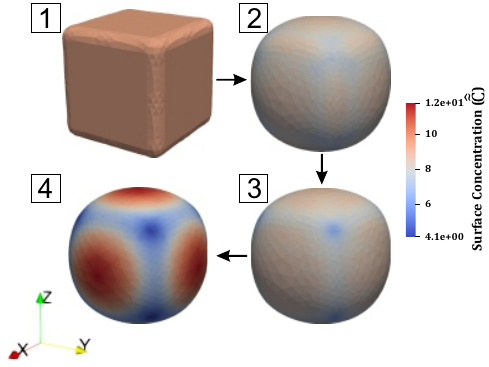}
    \caption{Temporal sequence of the surface concentration during the free contraction of cubes.}
    \label{fig:Fig4}
\end{figure*}

\begin{figure*}[ht]
    \centering
    \includegraphics[width=0.7\linewidth]{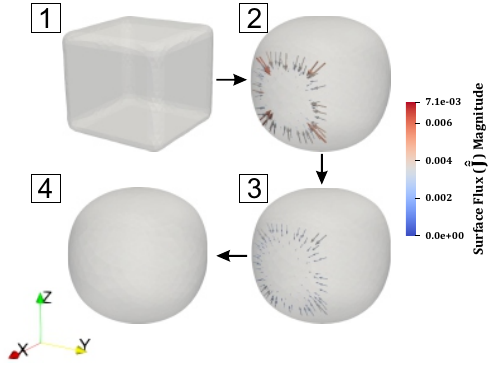}
    \caption{Temporal sequence of the surface flux during the free contraction of cubes.}
    \label{fig:Fig5}
\end{figure*}

\begin{figure*}[ht]
    \centering
    \includegraphics[width=1.0\linewidth]{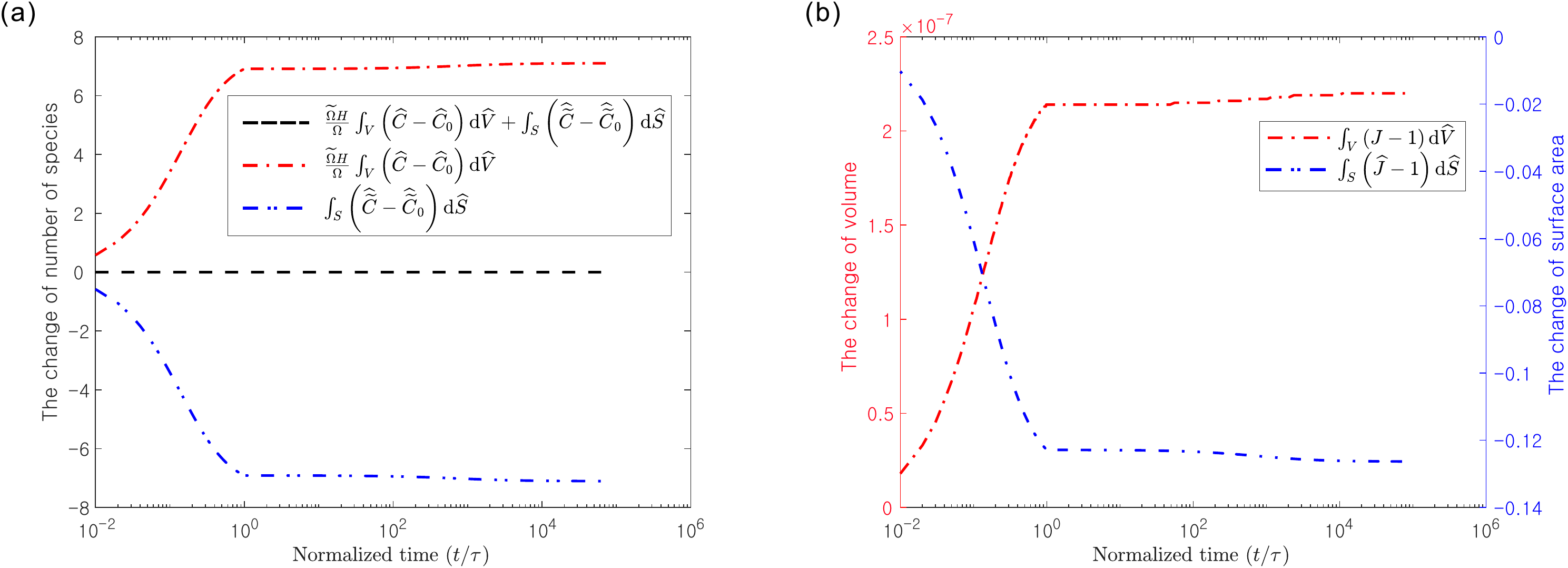}
    \caption{Surface energy drives the species migration between in bulk and on surface. (a) The numbers of species in bulk and on surface are tracked with respect to the normalized time. (b) The change of volume and surface area are tracked with respect to the normalized time. During the surface energy ramping ($\mathnormal{t}/\tau < 10^{0}$), the bulk is gaining the species, and the volume increases, while the surface is losing the species and the surface area decreases. While surface energy is being fixed ($\mathnormal{t}/\tau \geq 10^{0}$), the species still migrate from the surface into the bulk due to chemical potential. At equilibrium ($\mathnormal{t}/\tau \sim 1.6\times10^{5}$), there is no migration between the bulk and the surface. We note that the total number of species is constant at all times, and the volume increase of bulk is identical to the sum of the volume of species molecules moving into the bulk, i.e., $\int_{V}\left(\mathnormal{J}-1\right)\mathrm{d}\widehat{\mathnormal{V}}=\int_{V}\left(\widehat{\mathnormal{C}}-\widehat{\mathnormal{C}}_{0}\right)\mathrm{d}\widehat{\mathnormal{V}}$ due to the incompressibility condition of \Cref{eqn:MolecularIncompressiblity}.}
    \label{fig:Fig6}
\end{figure*}

\subsection{Free contraction of a cube with smooth edges}\label{sec:Free_contraction_of_a_sphere}
In this example, we first investigate the surface and bulk poroelastic effects on a   hydrogel cube. The cube is considered to be cast in a a fully swollen state, and then released in an environment in which it is free to deform, but with which it cannot exchange solvent ($\widetilde{\mathbf{J}}_{p}=0$). The two-stage solution procedure detailed above is followed here. 

We plot the temporal sequence of the finite element simulation for the chemical potential, bulk concentration, surface concentration, and surface flux in \Cref{fig:Fig2,fig:Fig3,fig:Fig4,fig:Fig5}, respectively, where all the values are normalized. We remove a quarter of the domain from the images of chemical potential in \Cref{fig:Fig2} and bulk concentration in \Cref{fig:Fig3} to show the contour plot on two sections of the interior along with the field on the surface. The surface flux in \Cref{fig:Fig5} is shown on only one face of the cube for visualization purposes. Images are taken at normalized time $\mathnormal{t}/\tau=0.0$, $1.0$, $8.4$, $7.8 \times 10^{4}$ in the clockwise, which are denoted by Step 1, 2, 3 and 4 in the following discussion. In Step 1 of \Cref{fig:Fig2,fig:Fig3,fig:Fig4,fig:Fig5}, the cubes are at the initial state by a homogeneous swelling stretch, $\lambda_{0}=3.215$. It is noteworthy that the cube is taken to have smooth edges (a case with sharp edges will be discussed in the following example). 

Step 2 of \Cref{fig:Fig2,fig:Fig3,fig:Fig4,fig:Fig5} corresponding to the time steps that the linearly ramping of the surface energy $\widehat{\widetilde{\gamma}}$ are included. This is for time $\mathnormal{t}/\tau=1.0$ which is significantly smaller (4 orders of magnitude) compared to the equilibration time. We note that the current interpolation scheme leads to unstable results if the ramping time is significantly shorter, pointing to the need for the development of a more advanced stabilization scheme in the future. This pathology might be a consequence of the three-field formulation. As the surface energy increases, the elastocapillary length $\widehat{\ell}$ becomes larger than the characteristic dimension $\mathnormal{H}$; as a result, the surface effects become dominant that the geometry becomes almost spherical. In Step 2 of \Cref{fig:Fig2}, the chemical potential is heterogeneous due to the deformation driven by surface energy. In Step 2 of \Cref{fig:Fig3,fig:Fig4}, the distribution of the bulk concentration on boundary is distinct from the surface concentration, and has not evolved as rapidly. This is owing to the fact that the species migration in bulk is slower than the one on surface, a specific modeling choice to highlight the lack of volumetric constraints for surface diffusion. However, in the equilibrium state, as shown in Step 4 of \Cref{fig:Fig3,fig:Fig4}, their distributions are similar on the surface.

On top of surface and bulk diffusion, in this system there is an exchange of species between the surface and the bulk, as captured in the strong form of the balance law for species conservation \Cref{eqn:ConservationLaw}. The species exchange between bulk and surface is easily tracked by accounting for the molecular incompressiblity condition of \Cref{eqn:MolecularIncompressiblity}, indicating that the change of volume of the bulk is the same as the total volume of the species migrating from bulk to surface. In \Cref{fig:Fig6}, we track the normalized total species population in bulk and on surface at all time steps by integrating the normalized concentrations over the bulk and the surface accordingly. We can observe the species migration between $\mathnormal{t}/\tau=0.0$ and $7.8 \times 10^{4}$ while the total number of species is conserved. In Step 2 of \Cref{fig:Fig4}, the surface concentration is high near the smoothened vertex of the cube because the deformation at that location is high compared to the facet, which subsequently leads to higher chemical potential and species migration. In Step 2 of \Cref{fig:Fig5}, the surface diffusion is mainly directed from the  edges to the interior of the facet. The direction of the surface diffusion is governed by the surface deformation and chemical potential (see \Cref{eqn:NormalizedFluxes}). 

Beyond Step 2, and for the rest of the transient response the surface energy ($\widehat{\widetilde{\gamma}}=1$) is maintained at the same level, and we exponentially ramp the time steps towards equilibrium. The deformation of the cubes is small beyond the initial surface energy ramping, but the concentration (especially in the bulk) significantly changes, driven by gradients of the chemical potential. In Step 3 of \Cref{fig:Fig3} and \Cref{fig:Fig4}, we can see the diffusion in bulk and on surface, and the species migrate toward the interior of the facet of the cubes. In Step 4 of \Cref{fig:Fig2}, the simulation reaches the equilibrium state, which is confirmed by the homogeneous chemical potential. the distribution of the bulk concentration on the boundary is similar to the surface concentration, as shown in Step 4 of \Cref{fig:Fig3} and \Cref{fig:Fig4}. In \Cref{fig:Fig6}, we can observe the significant migration between bulk and surface during ramping the surface energy ($\mathnormal{t}/\tau<1$), but the migration between bulk and surface is negligible with surface energy being fixed ($1<\mathnormal{t}/\tau \sim 7.8 \times 10^{4}$).

\begin{figure*}[ht]
    \centering
    \includegraphics[width=0.7\linewidth]{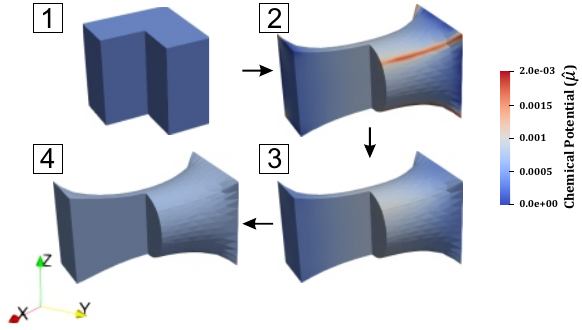}
    \caption{Temporal sequence of the chemical potential during the tension of cubes.}
    \label{fig:Fig7}
\end{figure*}

\begin{figure*}[ht]
    \centering
    \includegraphics[width=0.7\linewidth]{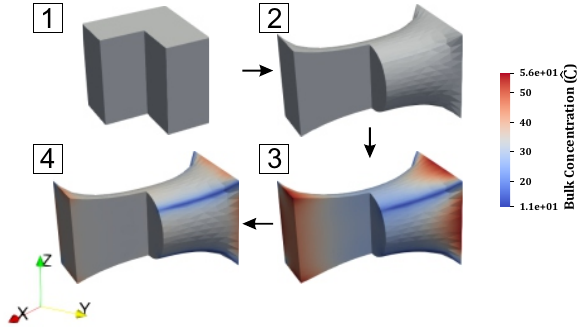}
    \caption{Temporal sequence of the bulk contraction during the tension of cubes.}
    \label{fig:Fig8}
\end{figure*}

\begin{figure*}[ht]
    \centering
    \includegraphics[width=0.7\linewidth]{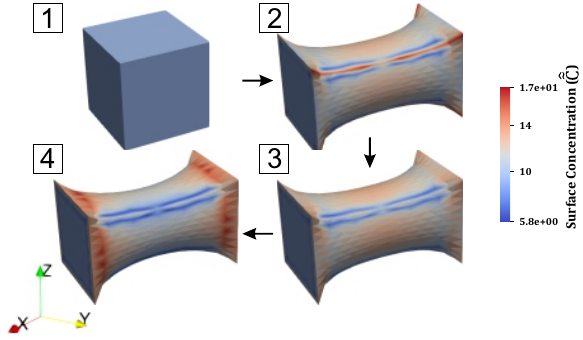}
    \caption{Temporal sequence of the surface concentration during the tension of cubes.}
    \label{fig:Fig9}
\end{figure*}

\begin{figure*}[ht]
    \centering
    \includegraphics[width=0.7\linewidth]{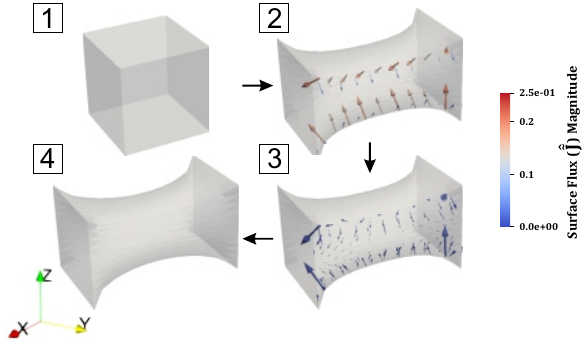}
    \caption{Temporal sequence of the surface flux during the tension of cubes.}
    \label{fig:Fig10}
\end{figure*}

\subsection{Uniaxial tension of a cube with sharp edges}\label{sec:Uniaxial_tension_of_a_cube}
In this example, we first investigate the surface and bulk poroelastic effects on a hydrogel cube subject to external load. Similar to the previous example the cube is considered to be cast in a a fully swollen state, and released  in an environment with which it cannot exchange solvent, but in this case it is subject to displacement controlled loading. The two-stage solution procedure detailed above is followed here as well. The lateral surfaces at $\mathnormal{x}(\mathnormal{t}=0)=\pm\mathnormal{H}/2$ are the clamped boundaries, and the others are free. In this case, the displacement boundary conditions at lateral surfaces, are ramped linearly from zero in the same timeframe that ramping of the surface energy takes place. 

In this case external mechanical loading along with surface and bulk energy contributions together dictate the transient response of the system. The other conditions are the same as in the previous example of \Cref{sec:Free_contraction_of_a_sphere}. We note that for this example, we have taken a cube geometry with sharp corners. We plot the temporal sequence of finite element simulations for the chemical potential, bulk concentration, surface concentration, and surface flux in \Cref{fig:Fig7,fig:Fig8,fig:Fig9,fig:Fig10}, respectively, where all the values are normalized. The surface flux in \Cref{fig:Fig10} is shown on only one face of the cube for visualization purposes. Images are taken at normalized time $\mathnormal{t}/\tau=0.0$, $1.0$, $5.4 \times 10^{2}$, $4.1 \times 10^{6}$ in the clockwise, which are denoted by Step 1, 2, 3 and 4 in the following discussion. 

In Step 1 of \Cref{fig:Fig7,fig:Fig8,fig:Fig9,fig:Fig10}, the cubes are at the initial state by a homogeneous swelling stretch, $\lambda_{0}=3.215$. In Step 2 of \Cref{fig:Fig7,fig:Fig8,fig:Fig9,fig:Fig10}, the cubes are deformed by the surface energy and stretch, after the initial load and surface energy ramping. The cubes are initially clamped at lateral sides, and we linearly ramp the surface energy $\widehat{\widetilde{\gamma}}=1$ and stretch $\varepsilon=68.9\%$ between $\mathnormal{t}/\tau=0.0$ and $1.0$. In Step 3 of \Cref{fig:Fig7,fig:Fig8,fig:Fig9,fig:Fig10}, we maintain the surface energy and the stretch, and we exponentially ramp the time-step size towards equilibrium. In Step 4 of \Cref{fig:Fig7,fig:Fig8,fig:Fig9,fig:Fig10}, the simulation approaches the equilibrium state. Through the second example, we can observe that surface concentration is distinct from the bulk concentration evaluated on the boundary.

Unlike the previous example in \Cref{sec:Free_contraction_of_a_sphere}, the cube at initial state has the sharp corners to investigate the impact of sharp geometric features to the numerical scheme. We can observe an oscillation of the surface concentration near the edges of the cube in \Cref{fig:Fig9}, which is partially regularized with time, reminiscent of the enforcement of an initial condition in a heat conduction problem. This originates from the weakly enforced continuity condition in \Cref{eqn:ConservationLawNeumannBC}, which states that the net surface flux on the boundary curve (between the adjacent surface elements) is always balanced. Of course, this can easily be resolved (as in the previous example), where  sharp features can be smoothened at the expense of computational cost owing to a finer non-structured mesh. Alternative, one could embark on deriving an appropriate stabilization scheme for this. We do not consider the impact of these initial oscillations to be significant to the transient response, similar to imposing discontinuous chemical potential boundary (with respect to the initial condition) for swelling of hydrogels \cite{bouklas2015nonlinear}. Again, this might be specific to the three-field formulation, and a two-field formulation could also alleviate the issue.

\section{Conclusion}\label{sec:Conclusion}
A continuum multiphysics formulation and corresponding finite element implementation has been presented to account for bulk and surface nonlinear poroelasticity in hydrogels. The governing equations for the response of a continuum body have been presented, and the general constitutive relations are derived in a thermodynamically-consistent manner, which is subsequently specialized for hydrogels. The proposed nonlinear theory is numerically solved through the mixed finite element method by employing the open-source finite element framework FEniCS. In addition, we provide FEniCS implementation details that could be important for problems where surface kinematics are of interest.

Two numerical examples are investigated to understand the effect of coupling bulk and surface poroelasticity. In the first example, where a cube with smoothened geometric features is investigated  (see \Cref{fig:Fig2,fig:Fig3,fig:Fig4,fig:Fig5,fig:Fig6}), we probe the variation of concentration, surface flux, and change of volume and surface area along with the chemical potential, which follow accounting for a fluid-like surface energy for the soft solid. As a result of the transient procedure and species exchange between bulk and surface, a smooth cube-like object is gradually transformed into an almost spherical one. From this observation, one may expect that the volume of the cube is decreased. Interestingly, the result is the opposite, as shown in \Cref{fig:Fig6}. It implies that the species collectively migrate from the surface into the  bulk because the volume increase of bulk is identical to the  volume of total species moving into the bulk; consequently, the volume of the cube rather increases in spite of the surface area decreasing. This result showcases that the multiphysical complication could play an important role in understanding the response of soft solids in length scales that surface effects dominate.

In the second example (see \Cref{fig:Fig7,fig:Fig8,fig:Fig9,fig:Fig10}), we further investigate the response of a cube with sharp features, under a external load, while still accounting for a fluid-like surface energy. Although the distributions of species concentration in bulk and on the surface are similar on the boundary in the first example (see \Cref{fig:Fig3,fig:Fig4}), the second example showcases that the pathways of species migrations can be significantly different (see \Cref{fig:Fig8,fig:Fig9}). In this example, we choose the same mobility constant in bulk and on the surface as the simplest case, but this result still provides an important insight into hydrogel-based applications at small scales. For example, one may use the distinct migration mechanism for the advanced design of hydrogel-based sensors and actuators. 

Even though we focus on hydrogels in this work, the general framework and finite element implementation developed here can allow studying several problems for solids as well in mechanobiology, where diffusion processes in the bulk and on the surface are coupled with the elastic response. One may adopt our numerical framework, but  different constitutive relations must be specialized for studying other materials. One specific example that the authors plan to pursue is the modeling of morphogenic processes in tissue mechanics and, more specifically modeling of contractile microtissues \cite{mailand2019surface,kim2020model,kim2023model}, where cells apply forces and contract the extracellular matrix (ECM), but at the same time can diffuse within the bulk of the  ECM but also on its periphery, having significant implications in problems like wound healing.  In this paper, we report numerical issues that can arise due to several reasons and propose approaches to alleviate them.  For the authors' future work, it would also be interesting to extend our model to consider growth mechanisms to account for surface morphology in evolving natural systems, such as encountered during tumor growth.

\section*{Acknowledgments}
JK and NB acknowledge the support by the National Science Foundation under grant no. CMMI-2129776. CYH acknowledges the support by the National Science Foundation under grant no. CMMI-1903308. 
FB thanks the project ``Numerical modeling of flows in porous media'' funded by Università Cattolica del Sacro Cuore, and the INDAM-GNCS project ``Metodi numerici per lo studio di strutture geometriche parametriche complesse'' (CUP\_E53C22001930001, PI Dr. Maria Strazzullo).

\section*{Declarations}
The authors declare that they have no known competing financial interests or personal relationships that could have appeared to influence the work reported in this paper.

\begin{appendices}
\section{Derivation of weak form of species conservation}\label{sec:Derivation of weak form of species conservation}
Staring form \Cref{eqn:ConservationLawBulk},
\begin{equation}\label{eqn:eqA1}
    \int_{\mathnormal{V}} \dot{\mathnormal{C}}\,\delta\mu \, \mathrm{d}V + \int_{\mathnormal{V}} \left(\boldsymbol{\nabla}_{\mathbf{X}}\cdot\mathbf{J}\right)\delta\mu \, \mathrm{d}V = 0
\end{equation}
By applying the product rule and divergence theorem \cite{holzapfel2000nonlinear}, and then substituting \Cref{eqn:ConservationLawSurf} into \Cref{eqn:eqA1},
\begin{equation}\label{eqn:eqA2}
    \int_{\mathnormal{V}} \dot{\mathnormal{C}}\,\delta\mu \, \mathrm{d}V - \int_{\mathnormal{V}} \mathbf{J}\cdot\boldsymbol{\nabla}_{\mathbf{X}}\delta\mu \, \mathrm{d}V 
    + \int_{\mathnormal{S}} \dot{\widetilde{\mathnormal{C}}}\,\delta\mu \, \mathrm{d}S + \int_{\mathnormal{S}} \left(\widetilde{\boldsymbol{\nabla}}_{\widetilde{\textbf{X}}}\cdot\widetilde{\mathbf{J}}\right)\delta\mu\, \mathrm{d}S = 0
\end{equation}
where the last term in \Cref{eqn:eqA2} can be rewritten by product rule and surface divergence theorem,
\begin{equation}\label{eqn:eqA3}
    \int_{\mathnormal{S}} \left(\widetilde{\boldsymbol{\nabla}}_{\widetilde{\textbf{X}}}\cdot\widetilde{\mathbf{J}}\right)\delta\mu\, \mathrm{d}S
    = - \int_{\mathnormal{S}} \widetilde{\mathbf{J}}\cdot\widetilde{\boldsymbol{\nabla}}_{\widetilde{\textbf{X}}}\delta\mu\, \mathrm{d}S 
    + \int_{\mathnormal{L}} \left(\widetilde{\mathbf{J}}\cdot\widetilde{\mathbf{N}}\right)\delta\mu \, \mathrm{d}L
\end{equation}
where the second term in the right-hand side of \Cref{eqn:eqA3} is assumed to be zero. By substituting \Cref{eqn:eqA3} into \Cref{eqn:eqA2},
\begin{equation}
    \int_{\mathnormal{V}} \dot{\mathnormal{C}}\,\delta\mu \, \mathrm{d}V - \int_{\mathnormal{V}} \mathbf{J}\cdot\boldsymbol{\nabla}_{\mathbf{X}}\delta\mu \, \mathrm{d}V 
    + \int_{\mathnormal{S}} \dot{\widetilde{\mathnormal{C}}}\,\delta\mu \, \mathrm{d}S - \int_{\mathnormal{S}} \widetilde{\mathbf{J}}\cdot\widetilde{\boldsymbol{\nabla}}_{\widetilde{\textbf{X}}}\delta\mu\, \mathrm{d}S = 0
\end{equation}
This equation corresponds to \Cref{eqn:WeakForm_ConservationLaw} in the manuscript.
\end{appendices}

\bibliographystyle{unsrtnat}
\bibliography{references}

\end{document}